# Tin (Sn) at high pressure: review, X-ray diffraction, DFT calculations, and Gibbs energy modeling

Guillaume Deffrennes[a*], Philippe Faure[a], François Bottin[b], Jean-Marc Joubert[c], Benoit Oudot[a*]

[a] CEA, DAM, VALDUC, F-21120 Is-sur-Tille, France

[b] CEA, DAM, DIF, F-91297, Arpajon, France

[c] Univ. Paris Est Creteil, CNRS, ICMPE, UMR 7182, 2 rue Henri Dunant, 94320 Thiais, France

* Corresponding authors:

Dr. Guillaume Deffrennes

Present postal address: National Institute for Materials Science, 1-1 Namiki, Tsukuba, Ibaraki 305-0044, Japan

e-mail: deffrennes.guillaume@nims.go.jp

Dr. Benoit Oudot

Postal address: CEA, DAM, VALDUC, F-21120 Is-sur-Tille, France

e-mail: benoit.oudot@cea.fr

# Abstract

An assessment of the Sn unary system is presented. First, the literature on phase equilibria, the thermodynamic properties, the volume and related properties, and shock compression of tin is thoroughly reviewed. Second, the Sn system is investigated by means of synchrotron X-ray diffraction in a diamond-anvil cell up to pressures and temperatures of 57 GPa and 730 K. New information is obtained on the thermal stability and thermal expansion coefficient of the γ ($I4/mmm$) and γ'' ($Im\bar{3}m$) phases. Third, density functional theory calculations are conducted on the six allotropic phases of tin observed in experiments using both a local density approximation (LDA) and a generalized gradient approximation (GGA) functional. This combined experimental and theoretical investigation provides further insights on the pronounced metastable nature of Sn in the 30 - 70 GPa range. Last, a Gibbs energy modeling is conducted using the recently proposed Joubert-Lu-Grover model which is compatible with the CALPHAD method. Special emphasis is placed on discussing extrapolations to high pressures and temperatures of the volume and of the thermodynamic properties. While the description of the heat capacity is approximate at moderate pressure, all available data are closely reproduced up to 2500 K, which is 5 times higher than the atmospheric pressure melting point of tin, and 150 GPa, which is almost 3 times the standard bulk modulus of β-Sn.

# 1. Introduction

Modeling of multi-component phase diagrams at high pressures is important for applications not only in geophysics, but also more and more in metallurgy [1–5]. Yet, to account for the effect of temperature, pressure, but also composition on the thermodynamic functions is a challenging task. On the one hand, the models should carry an explicit physical meaning, so that they can result in reasonable extrapolations under extreme conditions of $p$ and $T$. On the other hand, the treatment of solution phases imposes significant constraints. To enable extrapolations toward higher-order systems, it is required to describe end-members in mostly the entire $p$ - $T$ space, including in domains where they might be metastable or unstable. The potential spurious re-stabilization of end-members also has to be avoided, a typical case being solids re-stabilizing over the liquid at atmospheric pressure at high temperatures when the quasi-harmonic model is used to describe their heat capacity [6]. Physical Helmholtz energy models answer the first requirements, but a comprehensive and consensual framework to describe solution phases is not in place. They are not the focus of the present study. Phenomenological Gibbs energy models answer the second set of requirements within the framework provided by the CALPHAD method, but their application to high pressures has been mostly considered unsuccessful, as they tend to lead to unphysical predictions in the high $p$ - $T$ range [7]. The only alternative so far to build high-pressure Gibbs energy databases seems to be the model developed by Brosh *et al.* [6], which proposes a compromise to meet both the requirements mentioned above. This approach was successfully applied to the modeling of several binary systems [1,8–10].

In a recent study [7], the causes of the shortcomings of Gibbs energy approaches were discussed, and the recently proposed Joubert-Lu-Grover model [11,12] was identified as a promising candidate to extend CALPHAD databases toward high pressures. To evaluate the application potential of a Gibbs energy model, it should be applied to the modeling of a unary system, paying particular attention to extrapolations to high pressures and temperatures of the volume and of the thermodynamic properties. Then, as long as the models are compatible with the CALPHAD method, they can be extended to multi-component systems including solution phases.

The tin unary is an interesting system to put models to the test, because considerable experimental information is available up to relatively extreme temperatures and pressures. This element has a relatively low atmospheric pressure melting point compared with the more refractory materials commonly studied in geophysics, and data on the Sn liquidus are available up to 3000 K, which is 6 times higher than its $10^5$ Pa melting point. Besides, Sn is also relatively compressible, and static and dynamic compression studies were performed up to volume changes of 50% and 70%, respectively. The Sn unary system was modeled using Helmholtz energy approaches in several studies [13–16], but an explicit Gibbs energy modeling of this unary was never performed.

In the present study, a comprehensive review of the literature data on phase equilibria, thermodynamic properties, volume and related properties, and shock compression in the Sn system is provided. An experimental investigation is conducted by synchrotron X-ray diffraction (XRD) in a diamond-anvil cell (DAC) up to pressures and temperatures of 57 GPa and 730 K. Density functional theory (DFT) calculations are performed using both a local density approximation (LDA) and a generalized gradient approximation (GGA) functional. A thermodynamic modeling of the system is conducted using the recently proposed Joubert-Lu-Grover approach [11] that was so far only applied to the Pt-Os system [12].

## 2. Assessment of literature data

### 2.1 Solid phases

The known solid phases of the Sn unary system are presented in Table 1. The upper stability limit of the β-Sn phase at 298 K was determined as the mean of the 8 available measurements [17–24] with an expanded uncertainty with a 0.95 confidence level on the basis of a two-sided Student's t-distribution. The lower stability limit of the γ''-Sn phase at 298 K was determined in a similar fashion based on the 3 available measurements [22–24].

Table 1 - Stable solid phases of the Sn system, their crystal structure, and their stability limit along the 298 K isotherm and the $10^5$ Pa isobar. (a) Assessed data, (b) experimental data, and (c) modeled data. Data used to build the table are from [17–28]. TW stands for this work, DIA for diamond cubic, BCT for body-centered tetragonal, BCO for body-centered orthorhombic, BCC for body-centered cubic, and HCP for hexagonal close packed.

| Phase/ Lattice | Pearson Symbol/ Space Group/ Prototype | Stability at $10^5$ Pa (in K) | Stability at 298.15 K (in GPa) |
|---|---|---|---|
| α-Sn DIA | cF8 $Fd\bar{3}m$ C (Diamond) | 0 < T < 285±2[a], [27] <br> 0 < T < 286.3[c], [26] | - |
| β-Sn BCT | tI4 $I4_1/amd$ β-Sn | $T_{fus}$ = 505.078[a], [25] <br> 286.3[c] < T < 505.08[c], [26] | 0 < p < 10.2±1.9[a], TW <br> 0 < p < 9.7[c], TW |
| γ-Sn BCT | tI2 $I4/mmm$ In | - | 10.2±1.9[a] < p < 41.2±8.4[a], TW <br> 9.7[c] < p < 43.3[c], TW |
| γ'-Sn BCO | oI2 $Immm$ MoPt$_2$ | - | 30.5[b] < p < 70[b], lower limit: TW, upper limit: [24] |
| γ''-Sn BCC | cI2 $Im\bar{3}m$ W | - | 41.2±8.4[a] < p < 157[b], lower limit: TW, upper limit: [28] <br> 43.3[c] < p < 157[c], TW |
| δ-Sn HCP | hP2 $P6_3/mmc$ Mg | - | 157[b] < p < N/A, [28] <br> 157[c] < p, TW |

## 2.2 Phase equilibria

### 2.2.1 Phase equilibria at atmospheric pressure

The transition from β-Sn to α-Sn is commonly referred to as the tin pest, as it causes problems for numerous applications [29], notably because of the large volume increase of roughly 26% [30] associated with this transformation. Despite a long-standing interest [31], the equilibrium temperature of this transition and its underlying mechanisms are still questioned in the recent literature [26,32]. That is because the α to β transition is very sluggish [33–35], and long holding times of several hours are needed [33,36,37] to determine this thermodynamic equilibrium precisely. Furthermore, measurements are sensitive to the presence of ppm of impurities and to the material initial state [37]. Literature data on the thermal stability of α-Sn [27,31,32,34,36–43] are presented in Table 2. It appears from this review that differential thermal analysis (DTA) measurements [32,34,42,43] are scattered, and lead to higher transition temperatures than the former electrochemical [38,39] or dilatometric [31,36,37,40] studies. This is due to the fact that in these later DTA investigations, holding times are insufficient to reach a precise determination. In fact, these studies are more focused on the kinetics of the transformation, and the authors themselves considered the later measurement from Cohen *et al.* [36] as the actual equilibrium temperature. In the critical assessment from Gamsjäger *et al.* [27], the later measurement from Cohen *et al.* [36] was also considered to be the most reliable, along with the data from Raynor *et al.* [37]. Indeed, meticulous work is reported by Raynor *et al.* [37], and in both studies [36,37] the authors took care to obtain reactive α-Sn samples by thermal cycling. In the present work, the value of 285±2 K assessed by Gamsjäger *et al.* [27] is selected. It is in good agreement with the temperature of 286.35 K modeled by Khvan *et al.* [26].

Table 2 - Literature data on the α-Sn to β-Sn transition temperature at atmospheric pressure

| Ref | Method | T (K) |
|---|---|---|
| [38] Coh99 | Electrochemical analysis | 293 |
| [39] Coh08 | Electrochemical analysis | 291 |
| [40] Coh27 | Dilatometric analysis | 285 - 287.45 |
| [36] Coh35 | Dilatometric analysis | 286.35±0.1 |
| [37] Ray58 | Dilatometric analysis, 99.997% pure Sn, Fe as main impurity | 283.05 - 283.95 |
| [37] Ray58 | Dilatometric analysis, 99.997% pure Sn, Pb as main impurity | 286.15 - 286.75 |
| [41] Vnu84 | Change in electrical resistance | 304.8 |
| [31] Smi85 | Dilatometric analysis, commercial Sn | 286 |
| [31] Smi85 | Dilatometric analysis, zone-refined Sn | 285 |
| [42] Oji90 | Differential thermal analysis, heating rate of 5 K.min$^{-1}$ | 303 - 308 |
| [34] Gia09 | Differential thermal analysis, heating rate of 10 K.min$^{-1}$ | 326 |
| [43] Zuo13 | Differential thermal analysis, heating rate of 0.2 K.min$^{-1}$ | 297 |
| [32] Maz19 | Differential thermal analysis, heating rate of 2 K.min$^{-1}$ | 307.55 |
| [27] Gam12 | Critical assessment | 285±2 |

The melting point of the β-Sn is considered well-known [27], and is set at 505.078 K in the ITS-90 [25].

Finally, it is noted that an additional phase transition was suspected in the literature at roughly 440 K. First, an anomaly in the heat capacity was observed by Bartenev [44] at this temperature. Second, this hypothesis was further fuelled by Klemm and Niermann [45], as the authors reported on a small but sudden increase in the *a* parameter of the β-Sn tetragonal structure, along with a very slight discontinuity in enthalpy and thermal conductivity. However, this possible phase transition was question by Grønvold [46], as the heat capacity measurements from the authors showed no discontinuity. This trend was confirmed in numerous other datasets that are presented in the review provided by Khvan *et al.* [26]. Therefore, it is considered that there is no phase transition at 440 K.

## 2.2.2 Phase equilibria at high pressure

The available experimental studies [17–24,28,41,47–62] on high pressure phase equilibria in the tin system are reviewed in Table 3. Besides, the Sn phase diagram was also investigated by DFT calculations with LDA [63–66] and GGA [28,67–70] functionals, as well as by molecular dynamics simulations [71].

Table 3 - Experimental literature on high pressure phase equilibria in the Sn system

| Ref. | Method | $T$-range (K) | p-range (GPa) |
|---|---|---|---|
| [47] Dud60 | Change in electrical resistance along isobars in a tetrahedral-anvil apparatus | 531 - 771 | 0.6 - 10.7 |
| [17] Stag62 | Change in electrical resistance in a homemade piston-cylinder apparatus | 298 | 11.4 |
| [18] Str64 | Change in electrical resistance in gridle-anvil apparatus | 298 | 10.7 |
| [41] Vnu84 | Change in electrical resistance along isotherms | 273 - 304 | 0 - 0.07 |
| [48] Wei12 | Change in electrical resistance and in the power - $T$ curve in a resistive heating DAC | 1276 - 1996 | 10.4 - 44.7 |
| [49] McD62 | Thermal analysis | 506 - 529 | 0 - 0.5 |
| [50] Ken63 | DTA in a piston-cylinder apparatus | 518 - 637 | 0.5 - 4.9 |
| [51] Kin80 | DTA in a piston-cylinder apparatus | 505 - 630 | 0 - 3.7 |
| [52] Xu14 | Change in sound velocity (multi-anvil apparatus) | 538 - 694 | 0.7 - 4.6 |
| [53] Ram03 | Isochoric measurements and XRD in a DAC | 325 - 950 | 0.7 - 9.6 |
| [54] Bar63 | XRD in a tetragonal apparatus | 469 - 650 | 1.1 - 6.7 |
| [19] Jef66 | XRD in a tetrahedral apparatus | 298 | 9.2±0.3 |
| [20] Mii68 | XRD in an opposed anvil apparatus | 298 | 9.4±0.4 |
| [21] Oht77 | XRD in a multi-anvil apparatus | 298 | 9.31±5.2 |
| [22] Oli84 | XRD | 298 | 0 - 50 |
| [23] Liu86 | XRD in a DAC | 296±2 | 0 - 53.4 |
| [55] Des89 | XRD in a DAC | 298 | 52.1 - 120 |
| [56] Kie03 | XRD in a T-cup multi-anvil apparatus | 775 - 974 | 6 - 14 |
| [28] Sal11 | XRD in DAC | 298 | 148 - 194 |
| [57] Bri12 | XRD and change in the temperature - time profiles in a laser heated DAC | 1552 - 5520 | 18 - 105 |
| [24] Sal13 | XRD in DAC | 298 | 0 - 140 |
| [58] Bri17 | XRD and change in the power - temperature curve in a laser heated DAC | 1550 - 3100 | 22 - 105 |
| [59] Sch10 | Observation of laser induced speckle motion in DAC, or of textural change in quenched sample | 1324 - 2285 | 17.5 - 68 |
| [60] Mab99 | Shock compression measurements | 894 - 2295 | 5 - 49 |
| [61] Nik73 | Mössbauer spectrometry | 78 | 1.6 |
| [62] Bab62 | Change in volume | 505 - 533 | 0 - 0.9 |

Below 80 GPa, phase equilibria in the Sn unary system are difficult to characterize due to their pronounced metastable nature. The phases identified in various XRD studies [22–24,55] along the room temperature isotherm are presented in Fig. 1.

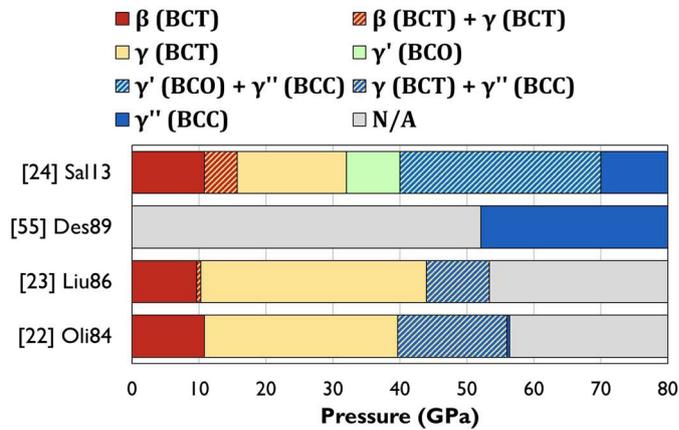

Fig. 1 – Phases identified in various XRD studies along the 298 K isotherm up to 80 GPa

With increasing pressure, β-Sn first transforms into an indium-like tetragonal structure noted γ. Both crystalline varieties were observed coexisting from 9.7±1 GPa to 11.1±8 GPa by Liu and Lui [23], and on a larger range in a later study [24].

Then, it was assumed in former studies that the γ-Sn phase would transform into a BCC structure noted γ'' in this work. Once again, both phases were observed coexisting from 40.5 to 56 GPa by Olijnyk and Holzapfel [22], and from 44±2 to at least 53.4 GPa in [23]. However, this metastable behavior was questioned by Desgreniers *et al.* [55], as only the γ'' phase was characterized by the authors at 52.1±0.2 GPa. Besides, in a more recent study, Salamat *et al.* [24] observed a transition at 32 GPa from the γ tetragonal variety to a BCO phase noted γ'. From this point, the transition with the γ'' BCC variety was noted from 40 GPa, but the γ' BCO phase was observed to exist up to 70 GPa. DFT calculations conducted by the authors [24] revealed that, above 30 GPa and up to at least 50 GPa, the potential-energy surface for the body-centered phases was flat and centered around b/a and c/a ratios of

1. Therefore, it appears that the BCC phase can transit into the BCT or BCO varieties by simple and continuous distortion of the lattice in response to slight deviations from hydrostatic conditions. This is why the indium-like BCT, the BCO and the BCC crystal varieties are all referred to the same γ letter designation in the present study.

Finally, a transition from the γ'' BCC phase to a HCP structure noted δ in this work was evidenced experimentally at 157 GPa at room temperature [28]. This finding was corroborated by several DFT studies [28,66,67,70].

The available phase diagram information is presented in Fig. 2. It is noted that the original data from Dudley *et al.* [47] were corrected. Instead of using the 1952 pressure scale from Bridgman [72], the former results from the author [73–75] that were eventually found to be more precise [76] were used to reprocess the measurements. Besides, regarding room temperature XRD studies where metastable two-phase equilibria were observed, it is noted that the equilibrium pressure was taken at the first occurrence of the higher-pressure phase.

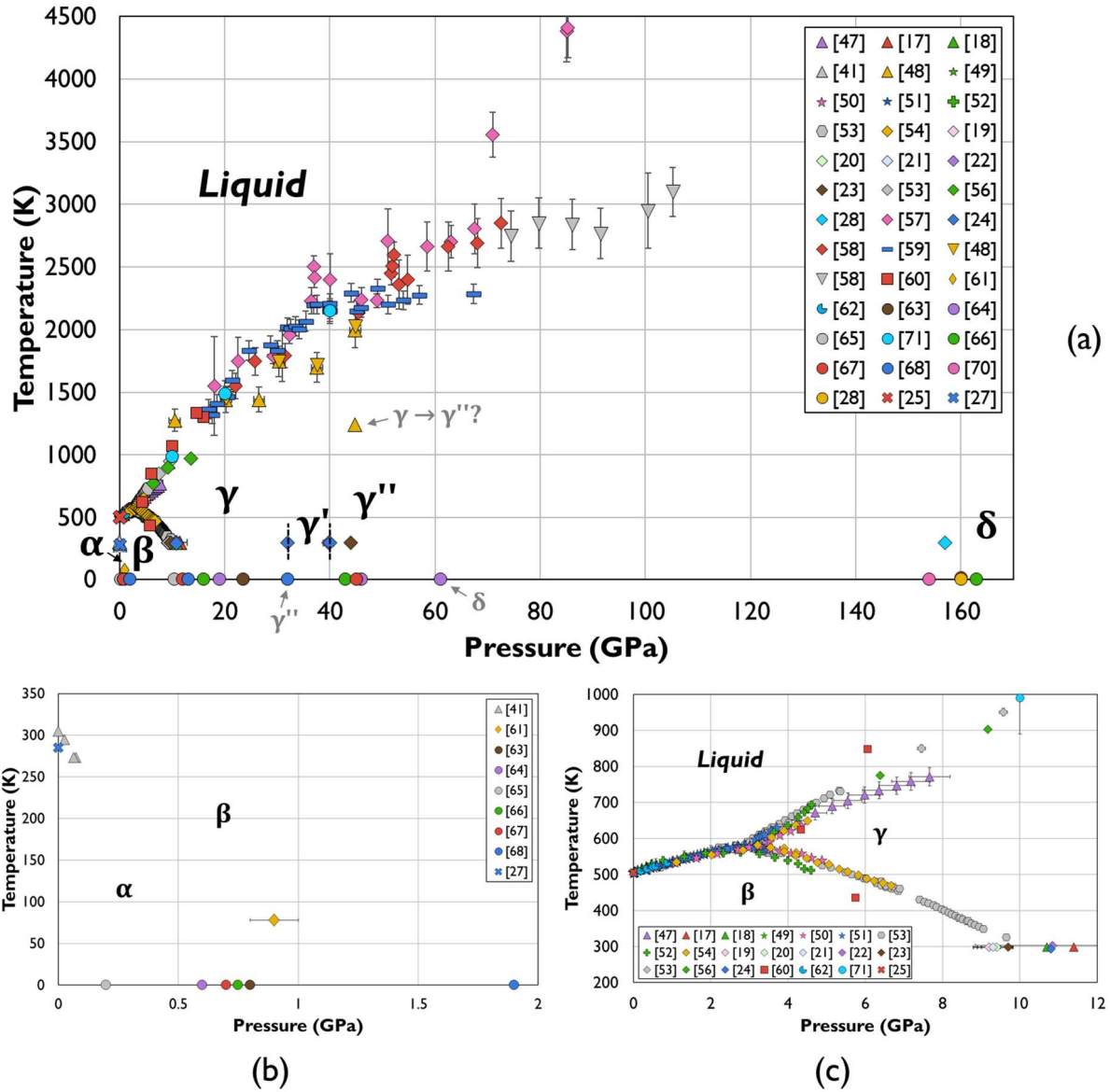

Fig. 2 – Experimental and *ab initio* data on the Sn phase diagram. (a) is an overview, (b) a focus on the α to β transition, and (c) a focus on the β-γ-liquid triple point. The symbols have the following meanings: △ change in electrical resistance, ☆ DTA, ✚ change in sound velocity, ⚬ isobaric method, ◇ XRD, ☐ visual determination, ▽ change in the laser heating profile, ☐ shock compression, ◊ Mössbauer spectrometry, ☉ change in volume, ○ *ab initio* calculations, ✖ assessment from the literature data.

In the low-pressure range, the available information on the α to β transition is presented in Fig. 2(b). The atmospheric pressure transition temperature measured by Vnuk *et al.* [41] is 20

K higher than the value assessed in [27] that was accepted in this work. The authors suggested that their measurements were too high due to a kinetic inhibition, but that the slope of the α-β boundary that they obtained was correct nonetheless.

Experimental data on the β-γ-liquid triple point are presented in Fig. 2(c). The liquidus of the β phase is very well-defined. Regarding the β-γ monovariant line, the results reported by Xu *et al.* [52] are shifted from roughly 30 K compared with the consistent measurements obtained in other studies [50,51,53,54]. Mabire *et al.* [60] observed an hysteresis of this transition under shock compression, that is delimited in Fig. 2(c) by the data point that seem to support the trend obtained in [52], and the one close to the liquidus of γ-Sn. The presence of this hysteresis under dynamic conditions was further evidenced by Briggs *et al.* [77]. It was suggested by Mabire *et al.* [60] that the true equilibrium transition was located somewhere in between their two data points. The β to γ transition at room temperature was studied in 8 independent studies, resulting in two slightly conflicting trends. 4 XRD studies [19–21,23] agree within their quoted uncertainties with a transition at roughly 9.4 GPa, whereas the other measurements [18,22,24] point toward a transition at 10.8 GPa, or even 11.4 GPa [17]. The data from Rambert *et al.* [53] on the β-γ monovariant line suggest a transition in between both the reported trend. Finally, the liquidus of the γ-Sn phase is the most disputed area. It was argued by Xu *et al.* [52] that Barnett *et al.* [54] underestimated the true equilibrium temperature in their XRD study because they detected the fast recrystallization temperature instead of the true melting point. The liquidus data point from Mabire [60] appears as an upper limit. Besides, the trend observed by Dudley *et al.* [47] is in poor agreement with the datasets available at higher temperatures. The measurements from Kingon and Clark [51] and Rambert *et al.* [53] seem the most reliable, because of their consistency with higher temperature measurements, as well as with several other datasets on the β-γ transition line and on the liquidus of the β-Sn phase.

The high-pressure liquidus of the γ-Sn phase is presented in Fig. 2(a). It was found by Briggs *et al.* [58] that the sudden increase that was observed around 70 GPa in a former study [57] was in fact due to the melting of the KBr pressure transmitting media. Then, the data from Weir *et al.* [48] are more scattered than other measurements, and a few of their liquidus data seem underestimated. Besides these two considerations, an overall satisfying agreement was reached between the available datasets [48,53,57,59,60], well within the uncertainties associated with this challenging experimental work, and this trend is further supported by molecular dynamics simulations [71].

**2.3 Thermodynamic properties**

**2.3.1 Thermodynamic properties at atmospheric pressure**

The literature on the heat capacity of the α, β and liquid phases was reviewed by Khvan *et al.* [26]. On this basis, a modeling was proposed by the authors. Another representation and a tabulation up to 300 K of the heat capacity data for both solid phases can be found in [78]. Regarding α-Sn, the description from [26] appears reliable up to at least 100 K. Above this temperature, the results are slightly more uncertain due to the lack of reliable experimental data. Regarding the β-Sn variety, the description proposed in [26] is very reliable up to the melting point of the phase, as consistent and abundant data were closely reproduced by the authors. The experimental information on the liquid phase are more disputed, but a consistent fit of several independent studies was reached by the authors up to at least 1100 K. Above this temperature, the heat content data measured by Feber *et al.* [79] up to 1800 K suggest that the heat capacity modeled in [26] starts to become underestimated.

The enthalpy change associated with each phase transition was also evaluated by Khvan *et al.* [26]. Regarding the α-Sn to β-Sn transition, a lack of data was noted by the authors, as only two conflicting experimental measurements [38,80] are available. An enthalpy of 2.101

kJ.mol$^{-1}$ was assessed by Khvan *et al.* [26], based on the later calorimetric determination of 2.23 kJ.mol$^{-1}$ from [80] and on their own DFT data. The enthalpy of fusion of tin is well-known and was assessed to be 7.179 kJ.mol$^{-1}$ [26].

**2.3.2 Thermodynamic properties at high pressure**

There is no experimental information available at high pressure on the thermodynamic properties of the phases. Several DFT studies have been conducted, and they will be discussed later along with the calculations performed in the present study.

**2.4 Volume and related properties**

**2.4.1 Volume and related properties at atmospheric pressure**

The volume, thermal expansion and bulk modulus data at atmospheric pressure on α-Sn and β-Sn were assessed in [30], and a tabulation can be found in [78]. Regarding α-Sn, few but consistent measurements are available from very low temperatures up to the decomposition temperature of the phase, and these data are closely reproduced in [30]. Regarding the β-Sn phase, more datasets are available, but they are also more conflicting. Nonetheless, the description proposed in [30] is supported by several consistent experimental studies.

Many studies are available regarding the density of the liquid phase. Several datasets were reviewed by Assael *et al.* [81], and a fit was proposed by the authors. Several additional research works were discussed by Alchagirov *et al.* [82]. The studies reported in both reviews [82–103] were compiled, and they are listed in Table 4. It is noted that several theses and reports cited in [82] could not be retrieved, and are thus not included here. The data will be presented later along with the results of the modeling.

Table 4 - Literature data on the density of liquid tin at atmospheric pressure

| Ref | Method | T-range (K) | Uncertainty (%) | Comment |
|---|---|---|---|---|
| [83] Kir52 | Archimedean | 504-2745 | - | Digitalized |
| [84] Kan68 | | 508-731 | 0.5 | |
| [85] Ber68 | | 523-1017 | 0.3 | Smoothed curve |
| [86] Ber70 | | 505-711 | 0.3 | Smoothed curve |
| [87] Kan73 | | 529-715 | - | |
| [88] She74 | | 500-1100 | - | Smoothed curve |
| [89] Bia86 | | 573-1492 | 0.4 | |
| [90] Wan03 | | 505-1133 | 0.05 | Digitalized |
| [91] Fis54 | Bubble pressure | 504-674 | - | Digitalized (low res.) |
| [92] Luc64 | | 565-954 | 0.2-0.7 | Digitalized |
| [93] Tim86 | | 800-1300 | 0.7 | Smoothed curve |
| [94] Fri97 | | 594-956 | - | Highly scattered |
| [95] Her60 | Pycnometric | 573-823 | 0.07 | Digitalized |
| [96] Thr68 | | 511-754 | 0.1 | |
| [97] Mat83 | | 507-614 | - | Smoothed curve |
| [82] Alc00 | | 506-761 | - | |
| [98] Yat72 | Large-drop technique | 510-924 | - | Digitalized |
| [99] Niz94 | | 485-1784 | - | |
| [100] Nak74 | Dilatometric method | 629-908 | 0.1 | Smoothed curve |
| [101] Dro79 | γ-ray attenuation | 512-1342 | - | Digitalized |
| [102] Nas95 | | 505-1503 | 0.75 | |
| [103] Sta06 | | 1100-1950 | 0.3 | Smoothed results |

The compressibility of liquid tin can be indirectly determined from measurements of sound velocity. First, the adiabatic bulk modulus $K_S$ can be calculated from the speed of sound $v_s$ and the density $\rho$ as follows:

$$K_S = \rho v_s^2 \tag{2.1}$$

Then, by definition $K_S$ is related to the isothermal bulk modulus $K_T$ as follows:

$$K_S = K_T(1 + \alpha \gamma T) \tag{2.2}$$

with $\alpha$ the volumetric thermal expansion coefficient, and $\gamma$ the Grüneisen parameter which can be expressed as:

$$\gamma = \frac{\alpha K_T}{\rho C_V} = \frac{\alpha K_s}{\rho C_p} \qquad (2.3)$$

with $C_V$ and $C_p$ the isochoric and isobaric heat capacities. Finally, it follows from Eq. (2.1), (2.2) and (2.3) that the isothermal bulk modulus can be calculated from the speed of sound as:

$$K_T = \frac{\rho v_s^2}{1 + \frac{\alpha^2 v_s^2 T}{C_p}} \qquad (2.4)$$

The available sound velocity measurements on liquid tin [104–113] were recently reviewed by Humrickhouse [114]. It was noted by the author that, if the measurements from Kleppa [104] and Berthou and Tougas [110] are outliers, the 8 other remaining datasets are consistent with each other. It appears from the present compilation that there was a conversion error in the treatment of the data from [110] by Humrickhouse [114]. This dataset is in fact in a much better agreement with the other measurements than suggested in the author's review. All the data will be presented afterwards when discussing the modeling.

**2.4.2 Volume and related properties at high pressure**

The volume of each phase listed in Table 1, except α-Sn, were investigated in various XRD studies [22–24,28,53,55] along the room temperature isotherm. A good agreement was reached, well within the quoted uncertainties. These data will be presented in a later time along with the results of the present experimental study.

The bulk modulus data on the β phase were reviewed in [7]. Measurements [115–118] are available along the room temperature isotherm up to 4 GPa. They are scattered within roughly 3 GPa, which represents a spread of approximately 5%. Nonetheless, the general trend is rather well defined, as it is supported by the available volume data.

Finally, the volume of the liquid phase was measured up to 4 GPa along the 573 K isotherm by Rambert *et al.* [53] based on an isochoric method. These data are the only experimental information available on the pressure dependence of the volume of the liquid phase.

**2.5 Shock compression**

For a material undergoing shock compression, the Rankine-Hugoniot equations are derived from the laws of mass, momentum and energy conservation along the shock front, leading to:

$$\frac{V_1}{V_0} = \frac{u_s - u_p}{u_s} \qquad (2.5)$$

$$p_1 - p_0 = \frac{u_s u_p}{V_0} \qquad (2.6)$$

$$H_1 - H_0 = \frac{1}{2}(p_1 - p_0)(V_1 + V_0) \qquad (2.7)$$

with $u_s$ the shock velocity, $u_p$ the particle velocity, $V$ the molar volume, $p$ the pressure, $H$ the enthalpy, and the subscript "0" and "1" referring to the state of the material before and after the shock, respectively. It is noted that Eq. (2.5), (2.6) and (2.7) are usually expressed in term of the density and the internal energy, but the present form is more adequate for Gibbs energy approaches.

It follows from the Rankine-Hugoniot equations that once $u_s$, $u_p$ and $V_0$ are known, information on the difference in enthalpy, volume and pressure between the initial state and the shocked state can be obtained. $u_s$ and $u_p$ were measured in various studies [119–125] along the principal Hugoniot of Sn, i.e. when the initial temperature is 298.15 K. Data are available up to 700 GPa, and are in a good agreement as it will be highlighted later on. Besides, experiments were also performed by Volkov and Sibilev [124] along the off-Hugoniot starting from 683 K at atmospheric pressure, where Sn is liquid.

It is noted that the 1958 study from Al'tshuler *et al.* [126] is not considered in the present study as the results are almost identical from the authors' 1962 compilation [121]. Furthermore, the Hugoniot data plotted on the phase diagram in [60] and [127] are also not considered, because they are model-dependent.

# 3. Methodology

## 3.1 Experimental investigation

High-pressure high-temperature X-ray diffraction experiments were carried out at the MARS beamline of the SOLEIL synchrotron [128] with an internal heated DAC which consisted in slight modifications of the ceramic DAC designed by LeToullec [129]. The angular aperture for the diffracted X-ray was increased up to 65° by using on the body part an Almax-Boehler geometry [130] diamond anvil (300μm culet size) glued on a conical WC seat. The DAC was also equipped with a membrane to allow for a fine tuning of the pressure.

A rhenium gasket with an initial thickness of 250 μm was pre-pressed to 40 μm and then drilled to form a hole of 170 μm in diameter. A 10 μm thickness tin sample (purity > 99.935% from Indonesian State Tin Corporation) was loaded in the high-pressure cavity (HPC) of the DAC together with several pressure-temperature markers: ruby balls ($Al_2O_3:Cr^{3+}$) and a 5 mole % samarium-doped strontium borate powder ($SrB_4O_7:Sm^{2+}$ referenced thereafter as borate) used as optical sensors [131] along with a 10 μm thickness platinum foil (purity > 99.95% from GoodFellow). The pressure transmitting medium, neon, was loaded at room temperature at the pressure of 2 kbar. Diffraction of neon was also used as a pressure marker when it was in the solid state. A photograph of the HPC is shown in Supplementary Note A.

The DAC was put inside a vacuum container in order to avoid graphitization of the diamonds at high temperatures (the pressure was kept inferior to 0.7 mbar during the experiment). Two K-type thermocouples were fixed close to the diamonds for the temperature regulation using a Watlow 988 device. We chose not to use the regulation of the thermocouple in the present experiment but let the temperature evolve at fixed power. A water-cooling circuit was also integrated in the container to cool the metallic body of the DAC. The monochromatic X-ray beam ($\lambda$=0.69266 Å – Zirconium K-edge) was focused to 14x12μm$^2$ FWHM at the sample position using mirrors in the Kirkpatrick-Baez geometry [128]. The angle dispersive X-ray diffraction signals were collected using a MAR345 imaging plate system, located at a distance of 244.9 mm from the sample. The Fit2D software [132] was used to treat and integrate experimental diffraction images after having determined the diffraction geometry from the diffracted signal of a LaB6 standard powder. Fullprof [133] was used to analyze diffraction patterns.

The DAC was slightly moved during the X-ray exposition (exposure time ranging between 10 and 60 sec.) in order to obtain the diffraction of both Sn and Pt in the same image. In the first part of the experiment, luminescence of the ruby R1 and borate $^5D_0$–$^7F_0$ lines was measured after excitation with a Nd:YAG laser (532 nm), using a 1800 lines/mm spectrometer (Jobyn Yvon HR320) coupled with a 64x1024 pixels CCD captor (C5809 Hamamatsu). Assuming no pressure dependence of the temperature-induced line shift of the ruby R1 line in our experiment [131], we used the IPPS-Ruby2020 pressure scale [134] to describe the pressure-induced line shift at room temperature together with a 3$^{rd}$ order polynomial law (Eq. 2 in [131]) to describe the temperature-induced line shift of ruby. In the case of borate, we used Eqs. 8 and 9 in [131] to describe the pressure and temperature-induced $^5D_0$–$^7F_0$ line shift.

In reason of the high sensitivity with temperature of the ruby luminescence, we used the coupling of the optical sensors with the Pt and Ne equation of state [135,136] to determine the pressure $P_{HPC}$ and temperature $T_{HPC}$ conditions within the HPC. The relation between $T_{HPC}$ and the temperatures $T_{c1}$ and $T_{c2}$ recorded by the thermocouples was fit with a linear relation ($T_{HPC}(K) = 295 + 0.48*(T_c - 295)$, with $T_c$ being the average of $T_{c1}$ and $T_{c2}$) which shows that strong temperature gradients occurred between the thermocouples locus and the HPC (around 100 K for $T_{HPC}(K) = 400$ K). This calibration relation was used in the second part of the experiment to determine $T_{HPC}$ from the thermocouples records, the error being set to $\Delta T_{HPC} = |T_{c1}-T_{c2}|/2$, $P_{HPC}$ being calculated from the Ne and Pt equation of states.

## 3.2 DFT calculations

In the present work, we perform *ab initio* calculations using the ABINIT code [137,138] and the projector augmented wave (PAW) [139,140] formalism. The six allotropic phases of tin are considered in this study: α-Sn (Diamond, with 2 atoms in the unit cell), β-Sn (BCT, with 2 atoms in the unit cell), γ-Sn (BCT, with 1 atom in the unit cell), γ'-Sn (BCO, with 1 atom in the unit cell), γ''-Sn (BCC, with 1 atom in the unit cell) and δ-Sn (HCP, with 2 atoms in the unit cell). For each, we compute their ground state properties, at various pressures from 0 to 1500 GPa. The pressure step is adjusted to have a good precision on the phase stability: 10 GPa between 0 and 100 GPa, 20 GPa between 100 and 200 GPa, and 100 GPa thereafter.

Two PAW atomic data are generated using the ATOMPAW code [141,142], considering 14 electrons in the valence, a cutoff radius of 2.5 bohr leading to a cutoff energy equal to 700 eV and an exchange and correlation (XC) energy treated using the LDA Perdew-Wang (PW) [143] and the GGA Perdew–Burke–Ernzerhof (PBE) [144] functionals, respectively. A careful treatment of the electronic density integration has been achieved with the use of a (24 × 24 × 24) Monkhorst–Pack (MP) mesh [145], whatever the crystal structure. This fine resolution of the MP mesh is needed in order to have well converged enthalpies [28].

For each target pressure, XC functional and crystallographic phase, the unit cell has been relaxed by conserving the crystal symmetries (Table 1) and by using a convergence criterion fixed to 0.2 meV between two consecutive iterations. To summarize, all the convergences (cutoff energy, MP mesh and structure relaxation) have been carefully checked in order to have an accuracy on enthalpies at most equal to 1 meV.

## 3.3 Thermodynamic modeling

### 3.3.1 Modeling of the Gibbs energy at atmospheric pressure

In the nomenclature used thereafter, the subscript "0" will be referring to the reference temperature of 0 K, and superscript "0" to the atmospheric pressure ($10^5$ Pa).

The Gibbs energy at atmospheric pressure is described using a so-called 3rd generation CALPHAD model [146,147]. The isobaric heat capacity of the crystalline phases is described based on a multi-frequency Einstein model as in [26,148–150]:

$$C_p^0 = 3R \sum_i \alpha_i \left(\frac{\theta_i}{T}\right)^2 \frac{\exp\left(\frac{\theta_i}{T}\right)}{(\exp\left(\frac{\theta_i}{T}\right) - 1)^2} + aT + bT^2 \qquad (3.1)$$

with $R$ the gas constant, $\theta_i$ the Einstein temperature for the i$^{th}$ mode of vibration, $\alpha_i$ the corresponding pre-factor constrained so that the sum all $\alpha_i$ is equal to the phase stoichiometry, and $a$ and $b$ parameters to account for electronic and anharmonic contributions [151]. The following expression is then obtained for the Gibbs energy:

$$G^0 = E_0 + \frac{3}{2}R \sum_i \alpha_i \theta_i + 3RT \sum_i \alpha_i \ln\left(1 - \exp\left(-\frac{\theta_i}{T}\right)\right) - \frac{a}{2}T^2 - \frac{b}{6}T^3 \qquad (3.2)$$

with $G$ the Gibbs energy referred to the enthalpy for the element in its stable form at 298 K and $10^5$ Pa, and $E_0$ the cohesive energy at 0 K. In order to avoid the spurious re-stabilization of crystalline phases at very high temperatures, the following empirical Gibbs energy expression taken from [152] is used above their melting point:

$$G^0 = E_0 + H' + \frac{3}{2}R \sum_i \alpha_i \theta_i + 3RT \sum_i \alpha_i \ln\left(1 - \exp\left(-\frac{\theta_i}{T}\right)\right) - S'T + a'T \ln(T) - \frac{b'}{30}T^{-6}$$

$$- \frac{c'}{132}T^{-1} \qquad (3.3)$$

$a'$, $b'$ and $c'$ are parameters set to ensure that (i) the $C_p$ converges toward a constant value at high temperatures that is lower or equal to the heat capacity of the liquid phase, and (ii) the $C_p$ and its temperature derivative are continuous at the melting point. $H'$ and $S'$ are parameters set to ensure the continuity of the enthalpy and entropy functions at the junction. It is noted that this treatment is the same as the SGTE method of extrapolation [153] used in 2nd generation databases.

The liquid and amorphous phase was described using the two-state model [154,155]. The atmospheric pressure Gibbs energy of the amorphous phase, noted $G^{am0}$, can be expressed using Eq. (3.2). However, no $b$ parameter [152] and only a single Einstein frequency should be used. Then, the atmospheric pressure Gibbs energy of the liquid and amorphous phase, noted $G^{liq-\ 0}$, is modeled as follows:

$$G^{liq-am0} = G^{am0} - RT\ln\left(1 + \exp\left(-\frac{\Delta G_d}{RT}\right)\right) \qquad (3.4)$$

In Eq. (3.4), $\Delta G_d$ is the Gibbs energy difference between the liquid-like and the amorphous-like states, and is expressed as follows:

$$\Delta G_d = B + CT + DT\ln(T) \qquad (3.5)$$

Recommendations on starting values and on the optimization procedure for the $B$, $C$ and $D$ parameters or for the description of $G^{am0}$ can be found elsewhere [26,146–149,152].

### 3.3.2 Modeling of the volume

The volume is described using the revision by Joubert *et al.* [11] of the model proposed by Lu *et al.* [156]. The volume is expressed as follows:

$$V = -cEi^{-1}\left(Ei\left(-\frac{V^p}{c}\right) - \frac{1}{K_T^p}\exp\left(-\frac{V^p}{c}\right)(p - p^0)\right) \qquad (3.6)$$

with $V$ the molar volume, $c$ a constant, and $Ei(x) = \int_{-\infty}^{x} e^t/t\, dt$ the exponential integral function that can be calculated numerically from tabulations. In the original model proposed by Lu *et al.* [156], $K_T^p$ and $V^p$ are the atmospheric pressure bulk modulus and molar volume, but after the revision from [11], this is only true at low pressure. In this work, the atmospheric pressure bulk modulus and molar volume are described based on the multi-frequency Einstein model proposed in [30]. This model was adapted to the Joubert-Lu-Grover framework in [7], leading to the following expression for $K_T^p$:

$$K_T^p = \frac{1}{\chi_{T_0}^0 + F_{cut} X \sum_i \frac{\alpha_i}{\exp\left(\frac{\theta_i}{T}\right) - 1}} \tag{3.7}$$

with $\chi_{T_0}^0$ the isothermal compressibility at 0 K and $10^5$ Pa, and $X$ a constant that governs the temperature dependence of the bulk modulus. It is underlined that the $\theta_i$ and $\alpha_i$ parameters are shared in common with the thermodynamic models presented in Section 3.3.1. $F_{cut}$ is a pressure cutoff function that is expressed as:

$$F_{cut} = \exp\left(-\frac{p}{p_{CUT}}\right) \tag{3.8}$$

with $p_{CUT}$ a cutoff pressure that is significantly higher than $10^5$ Pa. Then, $V^p$ is expressed as follows:

$$V^p = V_0^0 \exp\left(\int_0^T \alpha^p dT\right) \tag{3.9}$$

In Eq. (3.9), $V_0^0$ is the molar volume at 0 K and $10^5$ Pa, and $\int_0^T \alpha^p dT$ is expressed as:

$$\int_0^T \alpha^p dT = \frac{3R}{V_0^0} \sum_i \gamma_{i0}^0 \alpha_i \left( \theta_i F_{cut}' \left( \frac{\chi_{T0}}{e^{\frac{\theta_i}{T}} - 1} + \frac{XF_{cut}}{2\left(e^{\frac{\theta_i}{T}} - 1\right)^2} \right) \right.$$

$$\left. + F_{cut} \left( \frac{AT^2}{2} \left( \chi_{T0} - \frac{XF_{cut}}{2} \right) + \frac{T^3}{3} \frac{AXF_{cut}}{\theta_i} \right) \right) \quad (3.10)$$

with $\gamma_{i0}^0$ the 0 K and $10^5$ Pa Gruneisen parameter associated with the $i^{th}$ Einstein mode of vibration, and $A$ a parameter which meaning is discussed in [30]. $F_{cut}'$ is a second pressure cutoff function expressed as:

$$F_{cut}' = \exp\left(-\frac{p}{p_{CUT}'}\right) \quad (3.11)$$

with $p_{CUT}'$ a cutoff pressure that is significantly higher than $10^5$ Pa. The role of $F_{cut}$ in Eq. (3.7) and (3.10) is to ensure that the temperature derivative of $\alpha$ and $K_T$ becomes equal to 0 at high pressure, whereas $F_{cut}'$ ensures that the thermal expansion itself tends to 0 in this domain.

### 3.3.3 Modeling of the effect of pressure on the Gibbs energy

The Gibbs energy should be calculated as the sum of the atmospheric contribution detailled in Section 3.3.1, and of the integral of the volume over the pressure. In this work however, the following expression is used instead:

$$G = G^0 + cK_T^p \left( \exp\left(\frac{V^p - V}{c}\right) - 1 \right) \quad (3.12)$$

Eq. (3.12) was derived by Lu *et al.* [156], but becomes inexact after the modification proposed in [11] due to the fact that $V^p$ and $K_T^p$ became pressure dependent. A more detailed discussion can be found in [7]. The motivations and consequences of using Eq. (3.12) will be discussed in Section 4.3.1.1.

### 3.3.4 Modeling procedure

The optimization of the model parameters and the calculations were performed using the Thermo-Calc software [157]. The α (diamond cubic), β (BCT), γ (BCT), γ'' (BCC), δ (HCP), and liquid and amorphous phases were modeled following the procedure (how parameters were adjusted, which data was selected…) detailed in Supplementary Note B. The γ' BCO phase was not modeled as it was considered to be metastable. This choice is justified as follows. Among the 4 XRD studies along the 298 K isotherm [22–24,55] presented in Fig. 1, the presence of a BCO phase was observed only by Salamat *et al.* [24]. This can be explained by the fact that there is no energy barrier between the body-centered phases in the 30-50 GPa range [24], as discussed in Section 2.2.2. In the present XRD study, the diffraction pattern of the BCO phase could hardly be distinguished from the one of the BCT phase in this moderate pressure range (Section 4.1). This is consistent with our DFT calculations showing that the BCO lattice converges toward the BCT lattice from 15 to 35 GPa, and toward the BCC lattice at higher pressure (Section 4.2).

# 4. Results and discussion

## 4.1 Experimental investigation

Tin was studied up to 57 GPa and 730 K and the data are presented in the $p$ - $T$ space in Fig. 3. Most of the data are spread along 2 isotherms (at room temperature and around 670 K).

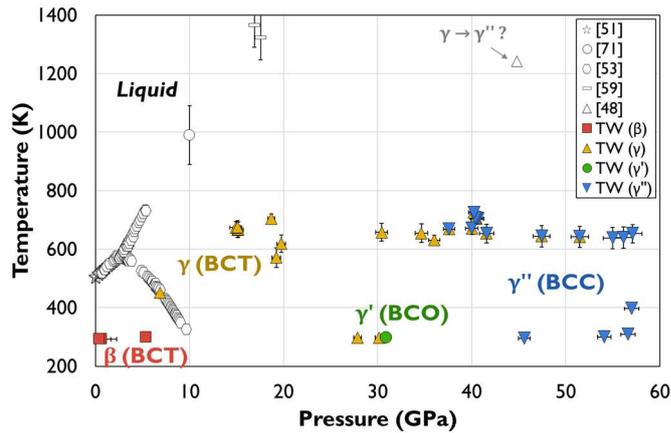

Fig. 3 – Results of the XRD investigation in the $p$ - $T$ space. Only a part of the literature data on phase equilibria that was presented in Fig. 2 is re-plotted for the sake of clarity.

In recent studies, the γ' BCO variety was observed coexisting with the γ'' BCC phase from 32 GPa to 70 GPa along the room temperature isotherm [24], and from 298 K up to at least 2130 K along the 46 GPa isobar [58]. In the present work, we cannot tell whether the BCO γ' structure is present above room temperature as its diffraction pattern would be too close to the one of γ-Sn to be distinguished with our data, even if we clearly observe a broadening of the (200) reflection of the BCT structure. Therefore, we chose to fit our diffraction patterns above room temperature only with the γ phase. The only pattern indexed with the BCO structure was recorded at room temperature at 31 GPa.

The γ to γ'' transition is found to start at 37.6 GPa at 669 K, and both phases are observed coexisting for more than 10 GPa or 50 K. The evolution of the diffraction patterns along the 650 K isotherm is presented in Fig. 4. In an ongoing study [158], this γ to γ'' transition was observed at roughly 26.4 GPa at 1050 K. These observations are consistent with the room temperature transition pressure of 41.2±8.4 GPa assessed in Table 1. They also suggest that the change in electrical resistance observed at 44.8 GPa and 1242 K by Weir *et al.* [48] was not caused by the γ to γ'' transition.

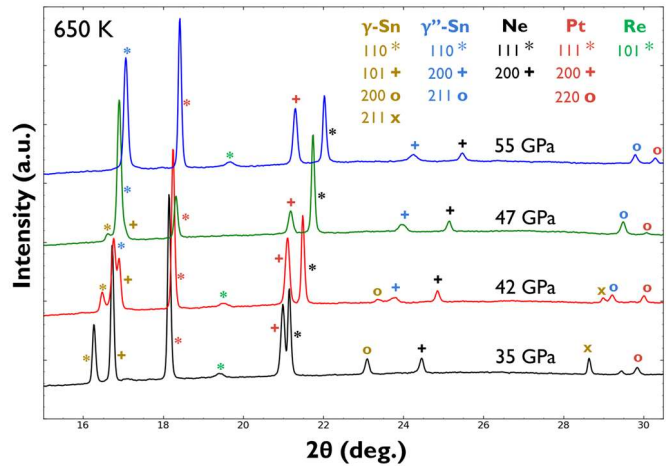

Fig. 4 – X-ray diffraction patterns showing the γ (BCT) to γ'' (BCC) transition along the 650 K isotherm

The volume data measured in this work along the room temperature isotherm are presented in Fig. 5(b-e). The obtained results are in good agreement with the literature data [22–24,28,53,55]. The results obtained at higher temperatures will be presented in Section 4.3.2.1.

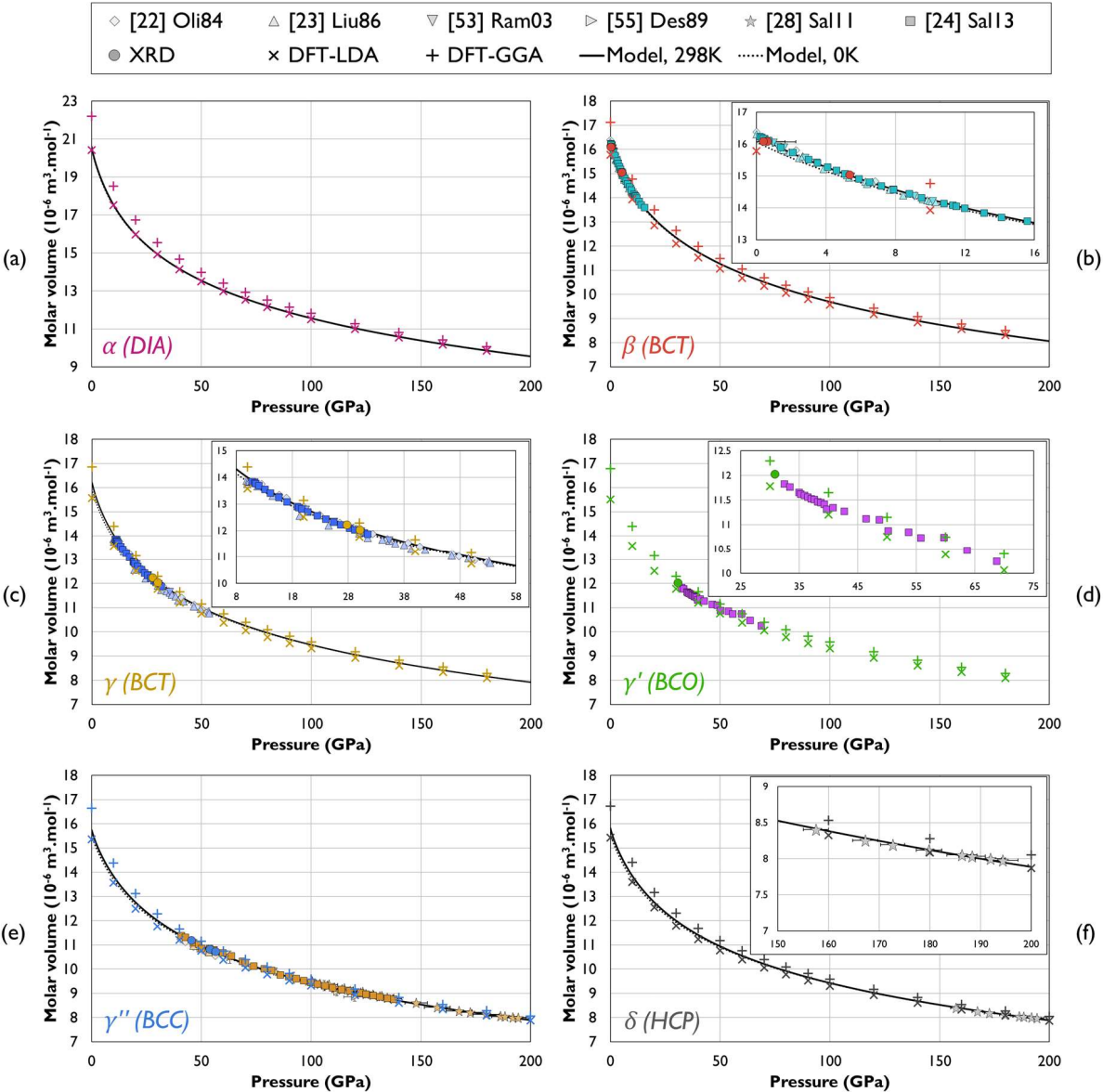

Fig. 5 – Volume of the crystalline phases of the Sn system as measured at room temperature in the present study and in the literature, calculated by DFT at 0 K, and modeled along the 0 K and the 298 K isotherms

## 4.2 DFT calculations

For each of the 6 allotropes considered, the volume calculated at 0 K is presented in Fig. 5. The results obtained with the LDA functional lead to an underestimated volume, whereas the DFT-GGA calculations lead to an overestimated volume compared with experiments. At higher pressures, the difference between both functionals is reduced, and a closer agreement with experiments is obtained, especially for the DFT-LDA calculations.

The difference in enthalpy with respect to the γ'' BCC phase is presented for each phase in Fig. 6(a-b). The calculated evolution with pressure of the lattice parameter ratios of the γ BCT and γ' BCO phases is shown in Fig. 6(c). For both the LDA and GGA calculations, the γ' BCO lattice converges toward the BCT lattice from 15 to 35 GPa, and then toward the γ'' BCC lattice hereafter. While the γ BCT variety is calculated to be slightly distorted from the BCC phase from 40 GPa to 160/180 GPa by LDA/GGA, both structures are extremely close in energy in this domain, with a difference of less than 2 meV. This result is consistent with the DFT-GGA calculations from [24], in which a flat potential-energy curve centered around the c/a ratio of 1 was obtained at 57 GPa, with a difference of less than 3 meV from c/a ratios of 0.94 to 1.02. Therefore, the present results further support the absence of a marked energy barrier between the γ BCT and γ'' BCC phases in the moderate pressure range. Last but not least, both the present LDA and GGA calculations point toward a re-stabilization of the γ'' BCC phase over the δ HCP variety at the extremely high pressure of 1350 GPa, as the volume of δ-Sn is predicted to become higher than the one the of γ'' phase from roughly 700 GPa.

The calculated phase transitions at 0 K are presented and compared with the results of the modeling in Table 5. It is noted that the δ HCP variety was calculated as the stable phase by DFT-GGA from 4.6 to 26.3 GPa. This result is not corroborated by experimental work, and is not shown in Table 5. For all the other phase transitions, the discrepancy between the DFT data and the model is somewhat consistent with the important spread observed between the different theoretical studies that can be seen in Fig. 2.

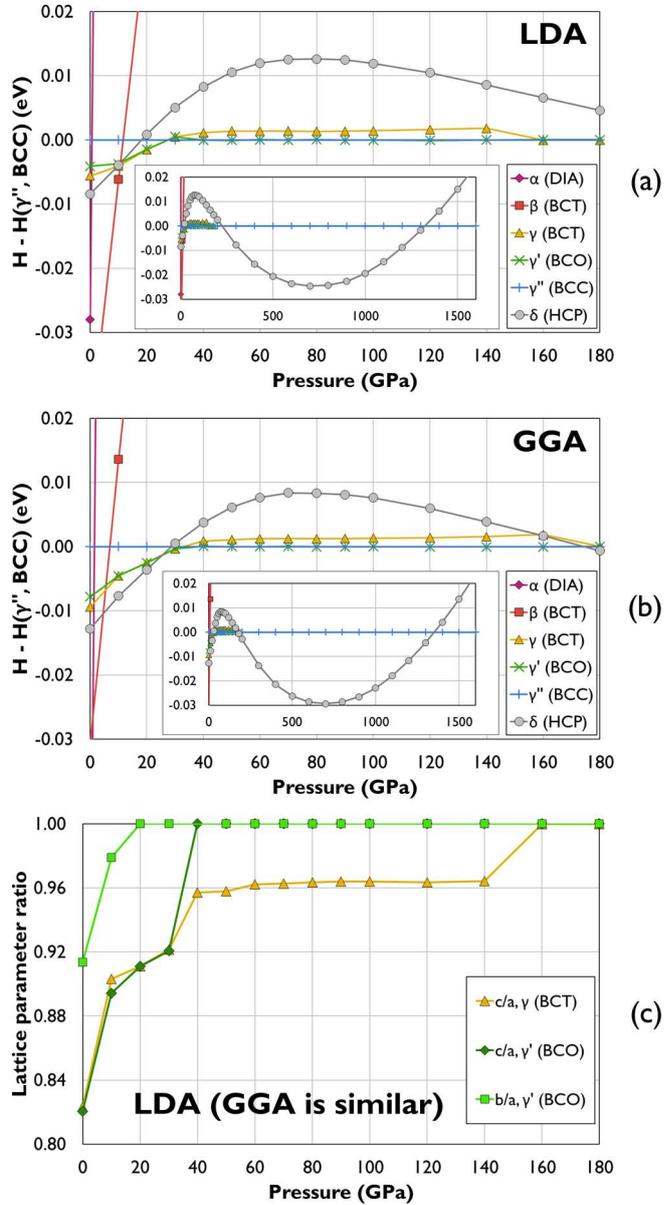

Fig. 6 – Difference in enthalpy with respect to the γ'' BCC phase calculated with a (a) LDA and (b) GGA functional. Inlets: data obtained at extremely high pressure for the γ'' and δ phases. (c) is the c/a and b/a lattice parameter ratio of the γ BCT and γ' BCO phases as calculated by DFT-LDA.

Table 5 - Phase transitions at 0 K obtained by LDA and GGA calculations compared with the results of the modeling, disregarding the presence of the γ' BCO phase and the stabilization predicted by DFT-GGA of the δ phase at low pressure

| Transition | P (GPa) (DFT-LDA) | P (GPa) (DFT-GGA) | P (GPa) (Model) |
|---|---|---|---|
| α → β | N/A | 0.9 | 0.3 |
| β → γ | 10.6 | 5.4 | 11.2 |
| γ → γ'' | 27.6 | 32.8 | 51.7 |
| γ'' → δ | 224.2 | 174.9 | 155.8 |
| δ → γ'' | 1323 | 1351 | N/A |

The volume of the crystalline phases of the system calculated at $10^5$ Pa by DFT-LDA and DFT-GGA are compared with the literature DFT data and with the results of the modeling in Supplementary Note C (Fig. S2). The same comparison is provided regarding the enthalpy of formation each phase in Fig. S3. Notable differences are observed between the different calculations, especially in regard to the enthalpy of formation.

**4.3 Thermodynamic modeling**

**4.3.1 Thermodynamic properties**

**4.3.1.1 Crystalline phases**

The heat capacity at atmospheric pressure of the phases modeled in this work is presented in Supplementary Note D (Fig. S4). A comparison between the model and the experiments for the α and β phases was provided in [26].

It follows from Eq. (2.2) and (2.3) that:

$$C_p = C_V(1 + \alpha\gamma T) \qquad (4.1)$$

At high pressure, the thermal expansion coefficient α tends to 0, and it follows from Eq. (4.1) that $C_p$ converges to $C_V$. At high pressure and temperature, 3R per mole of atoms is thus a reasonable limit for the isobaric heat capacity, as discussed by Brosh *et al.* [159]. In this work, an approximate expression, Eq. (3.12), is used to extend the atmospheric pressure Gibbs energy toward higher pressures. It follows from this approach that at high pressure, the $C_p$ increases back up to its atmospheric pressure value [7,11]. This behavior is illustrated in Supplementary Note D (Fig. S5), where the heat capacity of the phases is presented along the 1400 K isotherm. While this feature is unphysical, this approach enables to achieve compatibility with the atmospheric pressure CALPHAD databases built upon the SGTE method of extrapolation [153], as discussed in [7]. This method consists in setting the $C_p$ of crystalline phases to an arbitrary value above their atmospheric pressure melting point to avoid their spurious re-stabilization over the liquid at very high temperatures. In this work, this SGTE limit was set to *3R*, and thus, a reasonable heat capacity is obtained for crystalline phases under elevated conditions of *T* and *p*. However, their heat capacity will still go above this value below their $10^5$ Pa melting point, even at high pressure, and it will impact their entropy. Taking the practical case of β-Sn, and considering that the heat capacity should not rise above *3R* at high pressure, the present approach leads to an entropy overestimated by at least 4% at 505 K. Because variations of the Einstein temperatures with pressure are also not considered in the model, the overestimation of the entropy at high pressure is likely to be higher than that, but this value gives an order of magnitude. It is not straightforward to assess the impact of this approximate description of the entropy on the phase diagram. It is noted that this *3R* constraint was not applied to the liquid and amorphous phase, which heat capacity rises above this value at high *T* and *p*. This may lead to an overestimated thermal stability of the liquid phase in this domain, although the results of the present modeling rather suggest the opposite (Section 4.3.3).

In summary, the present approach leads to an inaccurate description of the heat capacity and entropy of the phases at high pressure. Nonetheless, this inaccuracy has arguably a limited impact on the phase diagram. It may be considered as an acceptable tradeoff, given that the model is readily compatible with the CALPHAD method, thus enabling the modeling of solution phases as well as extrapolations to multicomponent systems.

**4.3.1.2 Liquid phase**

The heat capacity at atmospheric pressure of the liquid and amorphous phase is presented in Fig. 7(a) along with the available data and the description from [26]. The same comparison is provided for heat contents in Fig. 7(b). The present description is in close agreement with the heat capacity measurements from Chen and Turnbull [160], whereas the modeling provided by Khvan *et al.* [26] is more consistent with the $C_p$ measured by Heffan [161]. Both descriptions satisfactorily reproduce most of the available heat capacity and heat content data, except for the more conflicting dataset from Klinkhardt [162], Yurchak and Philippov [163], and Wüst *et al.* [164]. As discussed in Supplementary Note B, the main difference between both assessments is that that a lower Einstein temperature was attributed to the amorphous phase in the present work, and that the high temperature heat content data from Feber *et al.* [79] were reproduced more closely, as highlighted in Fig. 7(b).

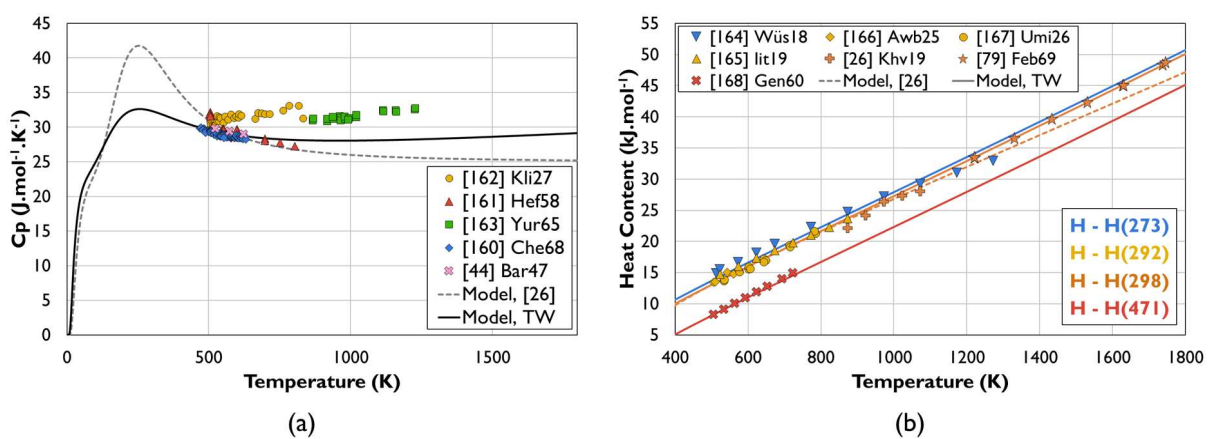

Fig. 7 – Atmospheric pressure (a) heat capacity and (b) heat content of the liquid and amorphous phase as modeled in the present work compared with experimental data [26,44,79,160–168] and the recent assessment from Khvan *et al.* [26]

## 4.3.2 Volume and related properties

### 4.3.2.1 Crystalline phases

The atmospheric pressure description of the volume and bulk modulus of α-Sn and β-Sn was accepted from [30]. For the other crystalline phases of the system, only the volume and bulk modulus at 0 K and $10^5$ Pa were adjusted. The parameters governing the temperature dependence of these properties were taken from the description of β-Sn due to the lack of data.

The volume of the crystalline phases calculated along the 0 K and 298 K isotherms is presented in Fig. 5. The available experimental data are well reproduced. Besides, the DFT-LDA data obtained in this work at 180 GPa are also closely reproduced.

The bulk modulus of β-Sn calculated along the room temperature isotherm is compared with the experimental data [115–118] in Supplementary Note D (Fig. S6). The later measurements from Bridgman [117] are closely reproduced.

The volume of the γ BCT and γ'' BCC phases calculated along different isotherms is presented and compared with the results of the present experimental study in Fig. 8. Regarding γ-Sn, a fair agreement between the model and the experimental data is observed, although the volume of the phase may be slightly overestimated. Regarding the γ''-Sn phase, the measurements are closely reproduced. These results along with those obtained on FCC Pt [11] and HCP Os [12] suggest that the Joubert-Lu-Grover model enables to achieve a satisfactory description of the volume up to temperatures of at least 1.5 times the melting point of the material of interest, and up to pressures of 2 to 3 times its standard bulk modulus.

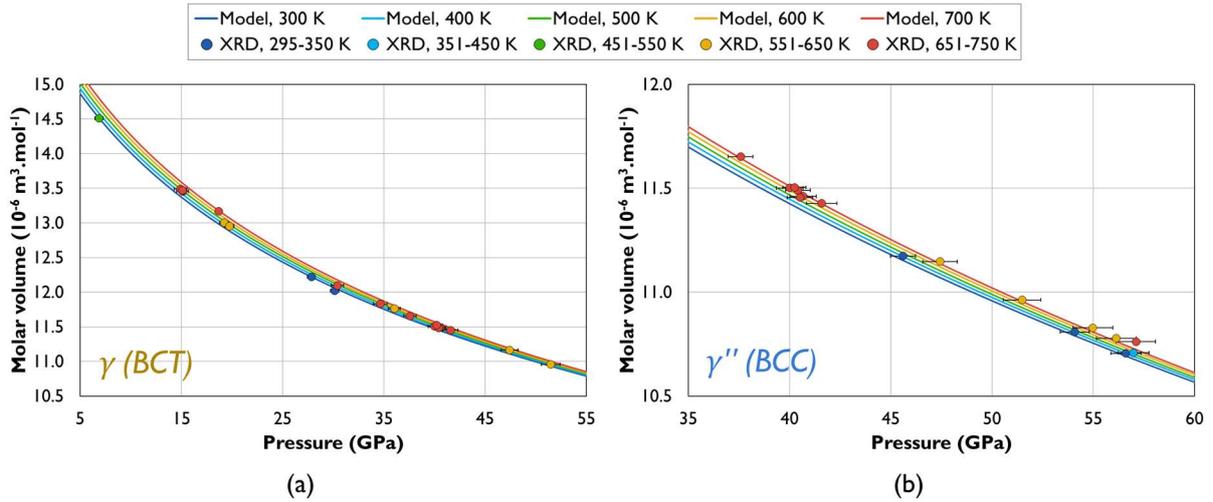

Fig. 8 – Modeled and measured molar volume of (a) γ-Sn and (b) γ''-Sn

**4.3.2.2 Liquid phase**

The volume of liquid Sn calculated at atmospheric pressure is presented along with experimental data [82–103] and the assessment from Assael *et al.* [81] in Fig. 9. The present description leads to a slightly lower volume than in [81] above 1000 K. This difference comes from the fact that additional datasets were considered in the present study, as well as from the weighting procedure as detailed in Supplementary Note B.

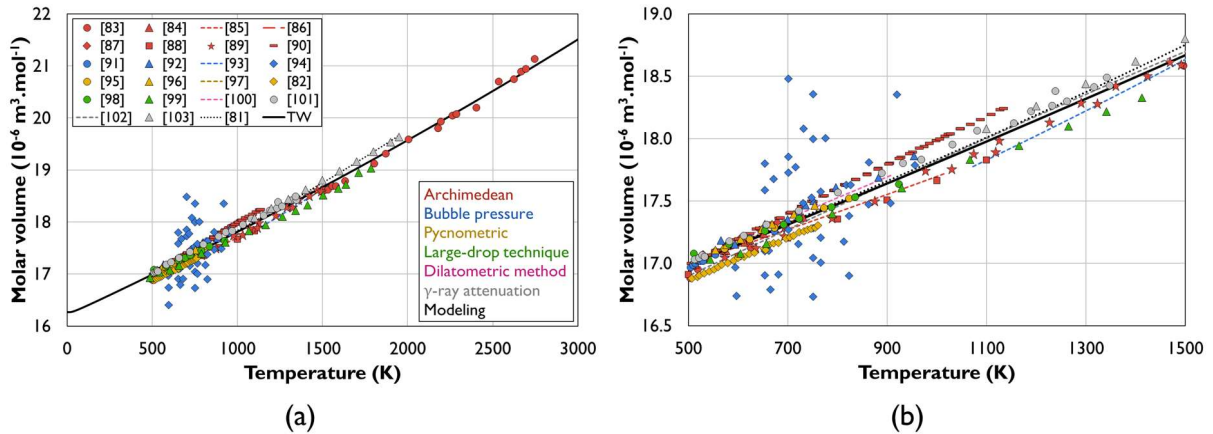

Fig. 9 – (a) Molar volume of liquid and amorphous Sn calculated at atmospheric pressure compared with experimental data and with a recent assessment from 0 to 3000 K. (b) is a magnified view in the 500 - 1500 K range, and the legend is the same as in (a).

The sound velocity in liquid Sn calculated at atmospheric pressure is compared with the available data [104–113] in Fig. 10(a). The bulk modulus of liquid Sn is calculated from the sound velocity based on Eq. (2.4) using the present description of its heat capacity and volume, and the results are presented in Fig. 10(b). The model and the data are in good agreement, except for the conflicting results from Gordon [105].

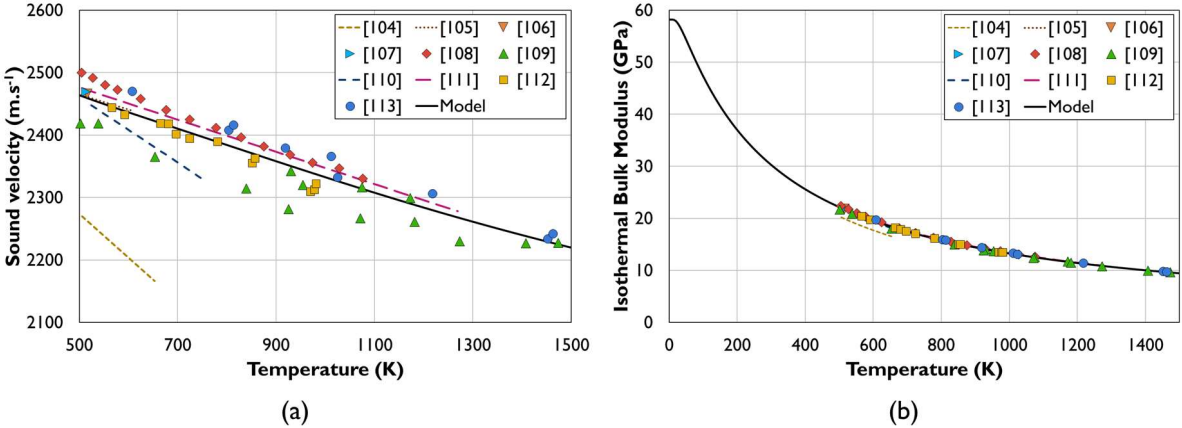

Fig. 10 – (a) Sound velocity in liquid Sn and (b) bulk modulus of the phase calculated at atmospheric pressure compared with the experimental data

The volume of liquid Sn calculated along the 505 K isotherm is presented in Fig. 11. A fair agreement was reached between the model and the measurements from [53], with a maximum discrepancy of 0.6% obtained at 1.5 GPa.

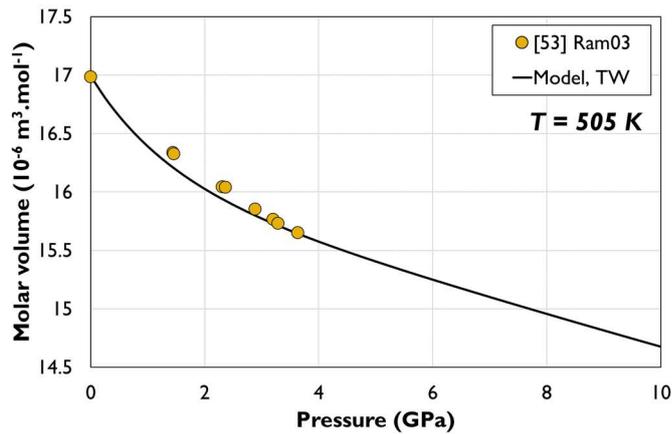

Fig. 11 – Molar volume of liquid Sn calculated along the 505 K isotherm compared with the experimental data

### 4.3.3 Phase equilibria and Hugoniot

The calculated Sn phase diagram is presented along with experimental data in Fig. 12. A comparison between the model and the data available along the principal Hugoniot and the 683 K off-Hugoniot is provided in Fig. 13. It can be seen from Fig. 12(a) that both these Hugoniot curves follow the liquidus monovariant line on a wide range of pressure. The phase proportion in this particular domain has a significant influence on the volume changes calculated along the Hugoniot, but it is not straightforward to compute it. Therefore, in Fig. 13(b) and (c), the upper and lower limit for the volume changes is plotted in these two-phase regions, instead of an arbitrary mean value. Thereafter, the results are discussed from low to high pressures.

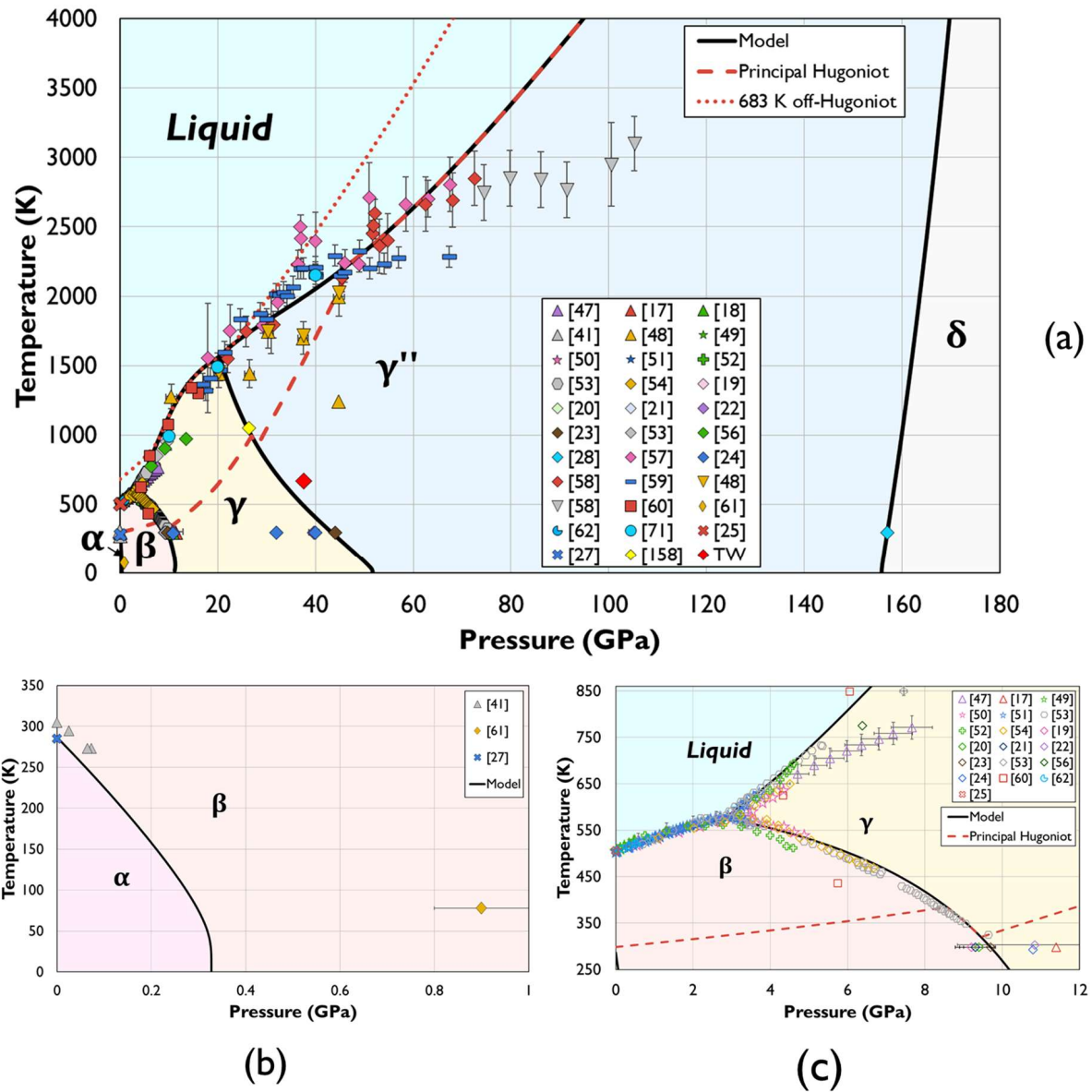

Fig. 12 – Calculated Sn phase diagram compared with the available data. (a) is an overview, (b) a magnified view of the low pressure α to β transition, and (c) magnified view of the β-γ-liquid triple point. The symbols are chosen based on the experimental method as detailed in the caption of Fig. 2.

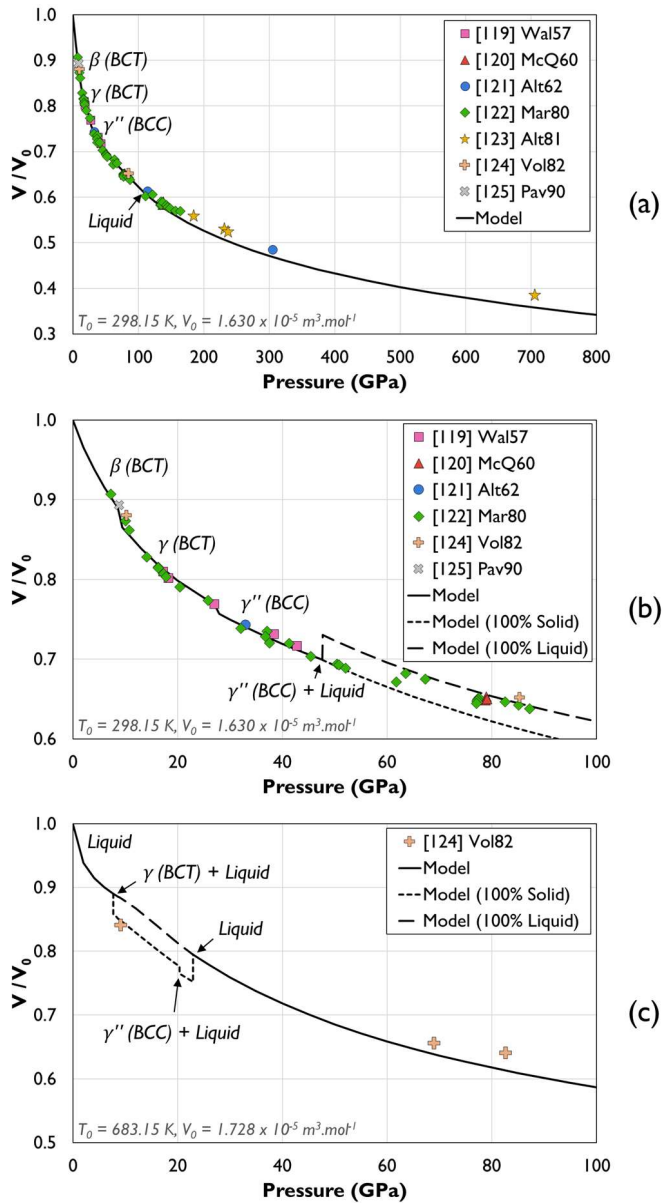

Fig. 13 – Calculated (a-b) principal Hugoniot and (c) 683 K off-Hugoniot compared with the experimental data. (a) is an overview of the principal Hugoniot, and (b) a magnified view in the 0 - 100 GPa range.

First, in the low-pressure range, it is shown in Fig. 12(b) that the calculated α to β monovariant line is in a satisfying agreement with the experimental data from Vnuk *et al.* [41], considering that the slope measured by the authors is correct, but that their data are shifted (Section 2.2.2).

Then, it is shown in Fig. 12(c) that satisfying results are obtained near the β-γ-liquid triple point, that is calculated at 577 K and 2.8 GPa. Regarding the liquidus of the β phase, the maximum deviation between the model and the data is of only 10 K. Regarding the more disputed liquidus of the γ phase and β-γ monovariant line, the measurements from Kingon and Clark [51] and Rambert et al. [53] are closely reproduced. The β to γ phase transition at room temperature is calculated at 9.7 GPa, which is in a better agreement with the 4 consistent studies [19–21,23] suggesting a slightly lower transition pressure.

When pressure is further increased, it is shown in Fig. 12(a) that a satisfying agreement is obtained between the model and the data near the γ-γ''-liquid triple point, that is calculated at 1566 K and 20.4 GPa. Regarding the transition from the γ BCT phase to the γ'' BCC variety, the results should be considered with care due to the pronounced metastable nature of Sn in this domain. At room temperature, the γ to γ'' transition is calculated at 43.3 GPa, compared with the pressure of 41.2±8.4 GPa assessed in this work based on the 3 available measurements [22–24]. At 669 K, this transition is calculated at 33.1 GPa, compared with the pressure of 37.6 GPa where both phases were first observed coexisting in the present experimental investigation. At 1050 K, the γ to γ'' transition calculated at 26.1 GPa is in a close agreement with the value of 26.4 GPa obtained in an on-going study [158].

Overall, a close agreement was reached between the modeled phase diagram and the available data in the 0 - 50 GPa and 0 - 2000 K range that was discussed so far. The consistency of the modeling in this domain is further supported by the fact the available shock compression data were closely reproduced, as shown in Fig. 13(b). This is especially true given that these data were not used in the optimization procedure of the crystalline phases, so this is an outcome.

The high-pressure high-temperature liquidus of the γ'' BCC phase is the most disputed region. A satisfying fit was obtained up to 70 GPa and 3000 K as shown in Fig. 12(a). Past this point, the data from [58] are not closely reproduced anymore, and the calculated liquidus temperatures appear overestimated. It would suggest that either the entropy of the liquid is underestimated compared to the one of the γ'' BCC phase, or its volume is overestimated. The first possibility seems unlikely in the present modeling, as discussed in Section 4.3.1.1. Regarding the second possibility, to set a lower volume for the liquid phase in this 60 - 150 GPa range would lead to a discrepancy between the calculated and measured volume changes along the Hugoniot curves. It is interesting to note that in previous modeling of the system based on Helmholtz energy approaches, significant deviations from the high-pressure liquidus data from Briggs *et al.* [58] were also obtained. At 80 GPa, the liquidus of the γ'' phase was measured by the authors at 2848±201 K. In the present work, the liquidus is calculated at 3384 K at this pressure, whereas temperatures of 5380 K, 3045 K, 3420 K and 3538 K were obtained by Molodets and Nabatov [13], Khishchenko [14], Cox and Christie [15] and Rehn *et al.* [16], respectively. It has to be noted however that the high-pressure liquidus is concave upward in this work, whereas concave downward liquidus were obtained in the mentioned Helmholtz energy modeling.

Next, the γ'' to δ transition is calculated to occur at 157 GPa at room temperature, in accordance with the only available experimental data from [28]. It is noted that the calculated γ''-δ monovariant line is completely arbitrary in the absence of further data.

Last but not least, it appears from Fig. 13(a) that there is an increasing gap between the calculated and the measured volume changes along the Hugoniot in the very high–pressure range from 100 GPa to 700 GPa. A similar trend was observed in [11] when applying the Joubert-Lu-Grover model to FCC Pt. In this model, the cold compression curve is described based on the empirical law discovered by Grover *et al.* [169], which is only valid up to pressures of roughly twice the standard bulk modulus of the material. For liquid Sn, it corresponds to approximately 120 GPa. Beyond this pressure, it can be deduced from [169] that the volume will start to become underestimated. This limitation appears clearly in Supplementary Note D (Fig. S7), in which the volume of γ''-Sn is calculated up to 1800 GPa and compared with the DFT calculations. Besides, it is also possible that the thermal expansion coefficient is underestimated at extremely high pressures and temperatures. This would be due to the mathematical form of the cut-off parameters introduced in the model, and to the fact that the rise from high temperatures of electronic or thermal vacancies contributions that are often assessed based on *ab initio* calculations [170] was not considered.

### 4.3.4 Model parameters

The model parameters to describe the atmospheric pressure Gibbs energy of the phases are presented in Table 6.

Table 6 - Model parameters for the description of the atmospheric pressure Gibbs energy of the phases, given in J.mol$^{-1}$

| Atmospheric pressure Gibbs energy of the crystalline phases below 505.08 K and of the amorphous phase based on Eq. (3.2) ||||||||
|---|---|---|---|---|---|---|---|
| $$G^0 = E_0 + \frac{3}{2}R\sum_i \alpha_i\theta_i + 3RT\sum_i \alpha_i \ln\left(1 - \exp\left(-\frac{\theta_i}{T}\right)\right) - \frac{a}{2}T^2 - \frac{b}{6}T^3$$ ||||||||
| Phase | $E_0$ | $\alpha_1$ | $\alpha_2$ | $\theta_1$ | $\theta_2$ | $a$ | $b$ | Ref |
| α-Sn | -9849.7499 | 0.67374 | 0.32626 | 218.4858 | 61.9652 | 1.1454E-03 | N/A | [26] |
| β-Sn | -7873.5615 | 0.64684 | 0.35316 | 159.075 | 61.122 | | | |
| γ-Sn | -4248.56 | | | 89.5 | | 3.9694E-03 | 1.5563E-05 | TW |
| γ''-Sn | 776.44 | 1 | N/A | 91.7 | N/A | | | |
| δ-Sn | 1937.92 | | | | | | | |
| Amorphous | -1941.79 | | | 80 | | 2.18705E-03 | N/A | |

| Atmospheric pressure Gibbs energy of the crystalline phases above 505.08 K according to Eq. (3.3) ||||||
|---|---|---|---|---|---|
| $$G^0 = E_0 + H' + \frac{3}{2}R\sum_i \alpha_i\theta_i + 3RT\sum_i \alpha_i \ln\left(1 - \exp\left(-\frac{\theta_i}{T}\right)\right) - S'T + a'T\ln(T) - \frac{b'}{30}T^{-6} - \frac{c'}{132}T^{-12}$$ ||||||
| Phase | $H'$ | $S'$ | $a'$ | $b'$ | $c'$ | Ref |
| α-Sn | 241.72132 | 0.73116428 | | 2.0809E+16 | -1.860263E+32 | [26] |
| β-Sn | | | | | | |
| γ-Sn | 2206.1005 | 5.63281 | N/A | -2.28095E+17 | -2.139945E+33 | TW |
| γ''-Sn | | | | | | |
| δ-Sn | | | | | | |

| Atmospheric pressure Gibbs energy of the liquid and amorphous phase according to Eq. (3.4) and (3.5) ||||
|---|---|---|---|
| $$G^{liq-am\,0} = G^{am\,0} - RT\ln\left(1 + \exp\left(-\frac{B + CT + DT\ln(T)}{RT}\right)\right)$$ ||||
| B | C | D | Ref |
| 5654.227 | -7.426965 | N/A | TW |

The model parameters to describe the volume and isothermal compressibility of the phases at are presented in Table 7.

Table 7 - Model parameters for the description of the molar volume and isothermal compressibility of the phases, given in m$^3$.mol$^{-1}$ and in Pa$^{-1}$ respectively

Atmospheric pressure molar volume and compressibility of the crystalline phases based on Eq. (3.7), (3.9) and (3.10)

$$V^0 = V_0^0 \exp\left(\frac{3R}{V_0^0} \sum_i \gamma_{i_0}^0 \alpha_i \left(\theta_i \left(\frac{\chi_{T_0}}{e^{\frac{\theta_i}{T}} - 1} + \frac{X}{2\left(e^{\frac{\theta_i}{T}} - 1\right)^2}\right) + \left(\frac{AT^2}{2}\left(\chi_{T_0} - \frac{X}{2}\right) + \frac{T^3}{3}\frac{AX}{\theta_i}\right)\right)\right)$$

$$\chi_T^0 = \chi_{T_0}^0 + X \sum_i \frac{\alpha_i}{\exp\left(\frac{\theta_i}{T}\right) - 1}$$

| Phase | $V_0^0$ | $\chi_{T_0}^0$ | $\gamma_{i_1}^0$ | $\gamma_{i_2}^0$ | $A$ | $X$ | Ref |
|---|---|---|---|---|---|---|---|
| α-Sn | 2.05065E-05 | 2.407E-11 | 1.6204 | -0.70538 | N/A | 2.8965E-12 | [30] |
| β-Sn | 1.60677E-5 | 1.6893E-11 | | 1.8394 | | | |
| γ-Sn | 1.59166E-05 | 2.02E-11 | 1.8394 | | 7.6494E-4 | 8.915E-13 | TW |
| γ''-Sn | 1.55055E-05 | 1.7E-11 | | N/A | | | |
| δ-Sn | 1.55742E-05 | 1.77E-11 | | | | | |

Atmospheric pressure molar volume and compressibility of the amorphous and liquid phase based on Eq. (3.7) and (S.1)

$$V^0 = V_0^0 \exp\left(\frac{3R}{V_0^0} Y \frac{\theta_E}{e^{\frac{\theta_E}{T}} - 1}\right), \quad \chi_T^0 = \chi_{T_0}^0 + X \sum_i \frac{\alpha_i}{\exp\left(\frac{\theta_i}{T}\right) - 1}$$

| $V_0^0$ | $\chi_{T_0}^0$ | $X$ | $Y$ | Ref |
|---|---|---|---|---|
| 1.62639E-05 | 1.7172E-11 | 4.833E-12 | 6.15856E-11 | TW |

Parameters from the Joubert-Lu-Grover model to extend the description toward high pressures using Eq. (3.6), (3.7), (3.10)

$$V = -c Ei^{-1}\left(Ei\left(-\frac{V^p}{c}\right) - \frac{1}{K_T^p}\exp\left(-\frac{V^p}{c}\right)(p - p^0)\right)$$

| Phase | $c$ | $p_{CUT}$ | $p_{CUT}'$ | Ref |
|---|---|---|---|---|
| α-Sn | 3.95E-06 | | 1.25E10 | |
| β-Sn | 3.26E-06 | | | |
| γ-Sn | 2.99E-06 | 1.5E9 | | TW |
| γ''-Sn | 3.055E-06 | | 3.5E10 | |
| δ-Sn | 3.034E-06 | | | |
| Liquid-amorphous | 2.97E-06 | 2.7E9 | 1.85E10 | |

## Conclusion

The Sn system was investigated by XRD up to 57 GPa and 730 K using a DAC, and new information on the volume, thermal expansion coefficient and thermal stability of the γ-Sn BCT and γ''-Sn BCC phases were obtained. This experimental study combined with DFT calculations further highlights the metastable nature of Sn in the 30 - 70 GPa range, that can be explained by the fact the γ BCT and γ' BCO phases are only slightly distorted and very close in energy from the γ'' BCC variety in this domain.

Based on the present investigation and on a thorough literature review, a thermodynamic modeling of the system was conducted. One of the aims of this work was to put to the test the recently proposed Joubert-Lu-Grover model. The approach taken here led to an approximate description of the heat capacity at moderate pressure, and as a result the sound velocity cannot be calculated accurately in this range. Nonetheless, the phase diagram, volume and Hugoniot data were reproduced closely up to at least 2500 K, which is 5 times higher than the atmospheric pressure melting point of Sn, and 150 GPa, which is almost 3 times the bulk modulus of β-Sn under standard conditions. At higher pressures, the Grover empirical law that is used to describe the cold compression curve becomes less precise, and the volume becomes underestimated. It is concluded that the present approach, which is readily compatible with the CALPHAD framework, appears promising to model multi-component phase diagrams at high pressure.

## Acknowledgements

This research did not receive any specific grant from funding agencies in the public, commercial, or not-for-profit sectors. Fruitful discussions within the French consortium in high temperature thermodynamics GDR CNRS n°3584 (TherMatHT) are acknowledged.

## Data availability

Our experimental data are given in a supplementary table which includes information for the pressure-temperature markers (cell-parameters for Pt and Ne and luminescence lines of ruby and borate). Our DFT data are provided in another supplementary table. A thermodynamic database (TDB) file made for the Thermo-Calc software [157] is given as supplementary materials.

# Supplementary Notes for:

# Tin (Sn) at high pressure: review, X-ray diffraction, DFT calculations, and Gibbs energy modeling

Guillaume Deffrennes[a*], Philippe Faure[a], François Bottin[b], Jean-Marc Joubert[c], Benoit Oudot[a*]

[a] CEA, DAM, VALDUC, F-21120 Is-sur-Tille, France

[b] CEA, DAM, DIF, F-91297, Arpajon, France

[c] Univ. Paris Est Creteil, CNRS, ICMPE, UMR 7182, 2 rue Henri Dunant, 94320 Thiais, France

* Corresponding authors:

Dr. Guillaume Deffrennes

Present postal address: National Institute for Materials Science, 1-1 Namiki, Tsukuba, Ibaraki 305-0044, Japan

e-mail : deffrennes.guillaume@nims.go.jp

Dr. Benoit Oudot

Postal address: CEA, DAM, VALDUC, F-21120 Is-sur-Tille, France

e-mail : benoit.oudot@cea.fr

**Supplementary Note A: Photograph of the high-pressure cavity**

An image of the high-pressure cavity (HPC) of the diamond-anvil cell (DAC) is shown in Fig. S1.

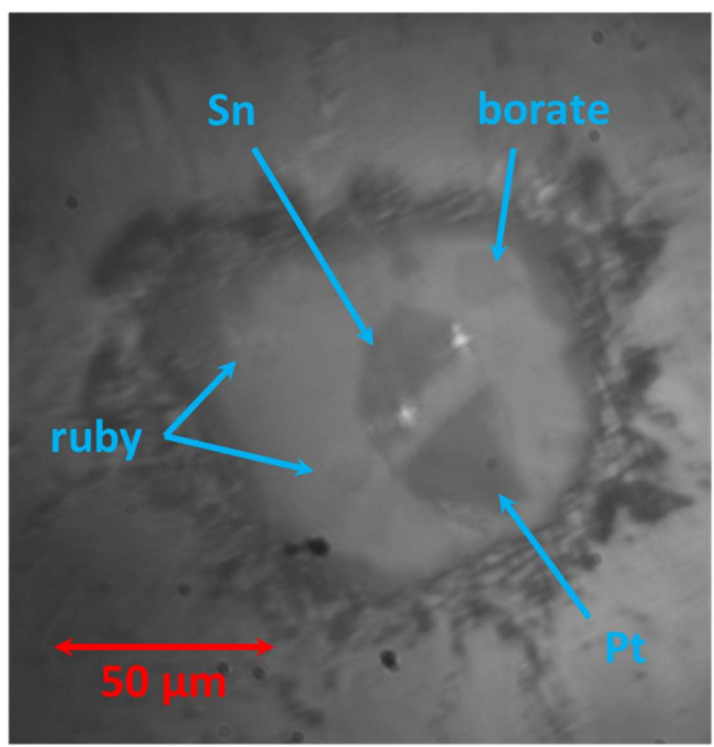

Fig. S1 – Image of the high-pressure cavity at 20 GPa and 650 K. The Sn and Pt samples are embedded in Ne and are surrounded by luminescence sensors (ruby and borate).

**Supplementary Note B: Detailed modeling procedure**

**Modeling of the α-Sn (diamond cubic) phase**

The atmospheric pressure descriptions of the Gibbs energy, and of the volume and bulk modulus of α-Sn were accepted from [26] and [30], respectively. The $c$ parameter of Eq. (3.6) that governs the pressure dependence of the volume was assessed based on the DFT calculations performed in the present work using the LDA functional. In the absence of data, the same $p_{CUT}$ and $p_{CUT}'$ parameters as for the β-Sn phase were accepted for α-Sn.

**Modeling of the β-Sn (BCT) phase**

The atmospheric pressure description of the volume and of the bulk modulus of β-Sn were accepted from [30]. The atmospheric pressure Gibbs energy description of the phase below its melting point was taken from [26]. Above the melting point of the phase, a change was made to the description proposed by the authors. In [26], the heat capacity of β-Sn tends towards the arbitrary value of 27.94 J.mol$^{-1}$.K$^{-1}$ at very high temperatures, whereas in this work, the value of 3R, i.e. 24.94 J.mol$^{-1}$.K$^{-1}$, was selected. That way, because Eq. (3.12) is used to compute the Gibbs energy function, the heat capacity of the phase is equal to 3R at high pressures and temperatures, which is considered a reasonable limit. The values for the $c$, $p_{CUT}$ and $p_{CUT}'$ parameters were constrained using the isothermal bulk modulus [115–118] and molar volume [22–24,53] measurements performed along the room temperature isotherm up to 4 and 15 GPa respectively, the volume calculated by DFT-LDA up to 180 GPa, and data on the liquidus of the phase and on the β-γ monovariant line.

**Modeling of the γ-Sn (BCT) phase**

The γ-Sn phase is metastable at atmospheric pressure, and there is no data on the temperature dependence of its heat capacity, volume and bulk modulus. Therefore, to model the atmospheric pressure Gibbs energy of the phase, the only parameters from Eq. (3.2) and (3.3) that were adjusted are $E_0$, and a single Einstein temperature $\theta_E$. All the other parameters were taken from the description of the β-Sn, which is the reference stable phase at 298.15 K. This treatment is consistent with the recent recommendation of Dinsdale *et al.* [147] to model lattice stabilities in the 3$^{rd}$ generation of CALPHAD databases. The atmospheric pressure volume and bulk modulus of the phase were modeled in a similar fashion. Besides from $\theta_E$ and from the $F_{cut}$ and $F_{cut}'$ functions that are equal to 1 at 10$^5$ Pa, the only parameters from Eq. (3.7) and (3.10) that were adjusted are $V_0^0$ and $\chi_{T_0}^0$, while the remaining parameters were accepted from the description of β-Sn. Finally, there is also not enough information to fit the $p_{CUT}$ parameter from Eq. (3.8) that is linked to the decrease of the temperature derivative of the thermal expansion coefficient and of the bulk modulus when pressure is increased. Once again, this parameter was thus set to the value selected for the β-Sn phase.

As a first step, the volume of the γ-Sn phase was modeled. To begin with, the $V_0^0$ parameter of the phase was set based on the volume obtained by DFT-LDA. These calculations lead to underestimated volumes: at 0 K and 10$^5$ Pa, the volume of α-Sn and β-Sn calculated by DFT-LDA are shifted by 10$^{-7}$ and 3x10$^{-7}$ m$^3$.mol$^{-1}$ respectively compared with the ones assessed based on experimental data in [30]. A comparable shift of 3.5x10$^{-7}$ m$^3$.mol$^{-1}$ was therefore applied to the DFT-LDA data for γ-Sn, and the $V_0^0$ parameter of the phase was set to the resulting value. Then, the $c$ and $\chi_{T_0}^0$ parameters of Eq. (3.6) and (3.7) were adjusted using the volume data measured by XRD along the room temperature isotherm in the literature [22–24,53] as well as in the present study. The upper limit of these measurements was reproduced because it led to a better agreement with the available phase equilibria data. To further constrain $c$ and $\chi_{T_0}^0$, the volume calculated by DFT-LDA at 180 GPa was taken as a lower limit for the modeling. Next, the $p_{cut}'$ parameter was adjusted using the high-temperature and high-pressure data on the volume of the phase obtained in the present study, under the constraint that it should be equal to the $p_{cut}'$ parameter of the γ''-Sn phase. That is because the γ BCT phase is only slightly distorted from the BCC variety in this pressure range, and this choice enabled to reduce the degrees of freedom in the modeling. In a second time, the $E_0$ and $\theta_E$ parameters of Eq. (3.2) and (3.3) were adjusted based on all the available information on the β-γ monovariant line. All the information reviewed in section 2.2.2 were considered in the process, except for the more conflicting datasets from Xu *et al.* [52], Stager *et al.* [17] and Mabire *et al.* [60].

**Modeling of the γ''-Sn (BCC) phase**

The modeling procedure for the γ'' BCC phase is very similar to the one described for the γ phase. Only 5 parameters were adjusted to describe the volume and of the thermodynamic properties of the phase, that are $E_0$, a single Einstein temperature $\theta_E$, $V_0^0$, $\chi_{T_0}^0$, and $p_{cut}'$. It was considered that not enough information was available to adjust the other parameters, and they were therefore taken from the description of the β-Sn.

First, the $V_0^0$ parameter of γ''-Sn was set to the DFT-LDA value calculated in this work after it was shifted by 1.5x10$^{-7}$ m$^3$.mol$^{-1}$. Then, the $c$ and $\chi_{T_0}^0$ parameters of the phase were adjusted to reproduce the volume data measured along the room temperature isotherm in the literature [22–24,28,55] and in the present study. Next, the $p_{cut}'$ parameter was adjusted based on the volume data measured at high temperature and high pressure in this work, under the constraint that this parameter was taken to be equal to the one used for the γ BCT phase.

Finally, the $E_0$ and $\theta_E$ parameters of the γ''-Sn phase were modeled based on the information on the transition from γ-Sn to γ''-Sn, and on the liquidus of the γ'' phase.

**Modeling of the δ-Sn (HCP) phase**

There is very limited information available on the HCP δ-Sn phase. Therefore, only 3 parameters, that are $E_0$, $V_0^0$, and $\chi_{T_0}^0$, were adjusted. For the $p_{cut}'$ parameter, the same value as obtained for the γ and γ'' phases was accepted. The same Einstein temperature as for the BCC γ'' phase was arbitrarily selected. All the other parameters were taken from the description of β-Sn.

The value for the $V_0^0$ parameter of δ-Sn was set based on the volume obtained by DFT-LDA that was shifted by 1.5x10$^{-7}$ m$^3$.mol$^{-1}$ as for the γ'' phase. The $c$ and $\chi_{T_0}^0$ parameters were adjusted based on the volumes measured along the room temperature isotherm by Salamat *et al.* [28]. Finally, the $E_0$ parameter was adjusted based on the γ'' to δ transition determined to occur at 298 K at 157 GPa by the same authors.

**Modeling of liquid Sn**

In the present work, modifications were made to the atmospheric pressure thermodynamic description proposed in [26]. The first reason behind this revision is that the Einstein temperature attributed to the amorphous phase by the authors was found to be higher than the one obtained for the high pressure γ and γ'' varieties in this work. As a result, the entropy of the liquid and amorphous phase was slightly lower than the one of these crystalline phases in the low temperature range, which is abnormal. The second reason is that the high temperature heat content data from Feber *et al.* [79] that suggest an increase in the heat capacity of the liquid phase from 1200 K were not closely reproduced by Khvan *et al.* [26], and this choice has a noticeable impact on the atmospheric pressure description of the bulk modulus obtained from sound velocity measurements. The Einstein temperature of the liquid and amorphous phase $\theta_E$ was set to the arbitrary value of 80 K, which is roughly 10 K lower than for the γ and γ'' crystalline phases. The $E_0$ and $a$ parameters of Eq. (3.2) and the $B$ and $C$ of Eq. (3.5) were adjusted based on the heat capacity data from Bartenev [44], Heffan [161] and Chen and Turnbull [160] that were selected by Khvan *et al.* [26]. All the available heat content data [26,79,165–168] were also considered in the optimization, except for the earlier measurements from Wüst *et al.* [164] which were found to be inconsistent with other studies. Besides, the melting point of 505.078 K established in the ITS-90 [25] and the enthalpy of fusion of 7187 J.mol$^{-1}$ selected by Khvan *et al.* [26] based on the measurements of Grønvold [46] were perfectly reproduced, as in [26].

Then, the atmospheric pressure molar volume and bulk modulus of the phase were modeled. It appears from the available density measurements that the volume of liquid Sn increases linearly with T up to at least 2750 K, suggesting a constant thermal expansion coefficient. The bulk modulus, however, decreases with T as suggested by the sound velocity measurements, and it follows from Eq. (3.10) that it leads to an increase in the thermal expansion coefficient. Therefore, it was found more suitable to simplify Eq. (3.10) as follows:

$$\int_0^T \alpha_{liq-am}^p \, dT = \frac{3R}{V_0^0} Y \frac{\theta_E}{e^{\frac{\theta_E}{T}} - 1} F_{cut}' \tag{S.1}$$

with $Y$ a constant that is linked to the product of the isothermal compressibility with the Grüneisen parameter. The $Y$ and $V_0^0$ parameters of Eq. (S.1) were obtained from all the available density datasets [82–103], except for the measurements from Friedrichs *et al.* [94] which are highly scattered. In the previous assessment of Assael *et al.* [81], the authors considered that the uncertainties for each dataset were of the same order of magnitude, and the data were only weighted according to the number of measurements. However, it appears that for each dataset except [94], the measurements are rather precise, because the spread is low, but they are not very accurate, because conflicting trends are reported. Therefore, in the present work, a weight proportional to the number of points was attributed to each dataset, so that each reported trend would have the same weight in the optimization. Next, the available sound velocity measurements [104–113] were converted into isothermal bulk modulus data based on Eq. (2.4), using the previously obtained heat capacity and volume description. On this basis, the $\chi_{T_0}^0$ and $X$ parameters of Eq. (3.7) were adjusted. In the process, each dataset was weighted following the same logic as discussed above for the volume, except for the measurements from Kleppa [104] which are clearly outliers and were therefore not considered in the optimization.

Finally, the parameters governing the pressure dependence of the volume were adjusted. First, the $c$ parameter was adjusted based on the shock data along the principal hugoniot available from 110 GPa to 200 GPa [121–123]. What was considered a reasonable lower limit for the volume changes was reproduced as it was found to be more consistent with the available data on the liquidus of γ''-Sn. Then, the $p_{cut}'$ parameter was assessed using all the information available reviewed on Section 2.2.2 on the high-pressure liquidus of the γ'' phase, expect for the measurements from Weir *et al.* [48] and for the data above 70 GPa from the former study from Briggs *et al.* [57]. Last, the $p_{cut}$ parameter was assessed using the data on the liquidus of the γ Sn obtained up to 6 GPa by Kingon and Clark [51] and Rambert *et al.* [53]

**Supplementary Note C: Volume and enthalpy of formation calculated by DFT compared with literature DFT data and with the results of the modeling**

The volume of 6 allotropic phases of tin calculated at $10^5$ Pa by DFT-LDA and DFT-GGA is compared with the literature DFT data and with the results of the modeling in Fig. S2. The same comparison is provided regarding the enthalpy of formation of each phase in Fig. S3.

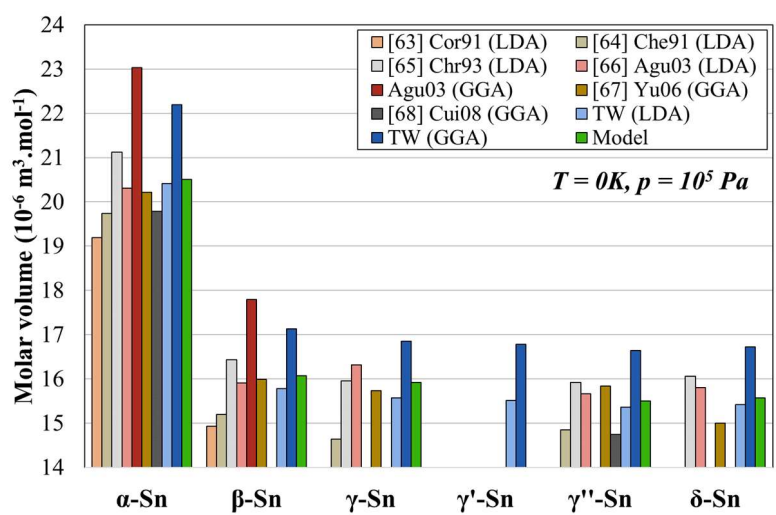

Fig. S2 –Volume of the crystalline phases calculated by DFT-LDA and DFT-GGA compared with literature DFT data and with the results of the modeling

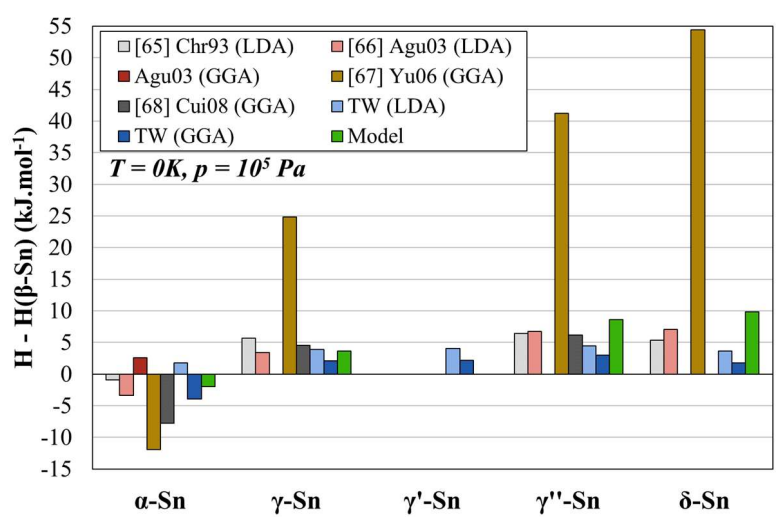

Fig. S3 – Enthalpy of formation of the crystalline phases with respect to β-Sn calculated by DFT-LDA and DFT-GGA compared with literature DFT data and the results of the modeling

**Supplementary Note D: Additional modeling results**

In this note, additional figures are presented to further show the agreement between the model and experimental data, and to highlight the limitations of the model.

The heat capacity of the α, β, γ, γ', γ'', δ and liquid and amorphous Sn phases are presented along the atmospheric pressure isobar in Fig. S4, and along the 1400 K isotherm in Fig. S5.

A comparison between the model and the experimental data on the bulk modulus of β-Sn available along the room temperature isotherm is presented in Fig. S6.

The volume of the γ'' BCC phase calculated along the 0 K isotherm is compared with DFT calculations up to the extremely high pressure of 1800 GPa in Fig. S7.

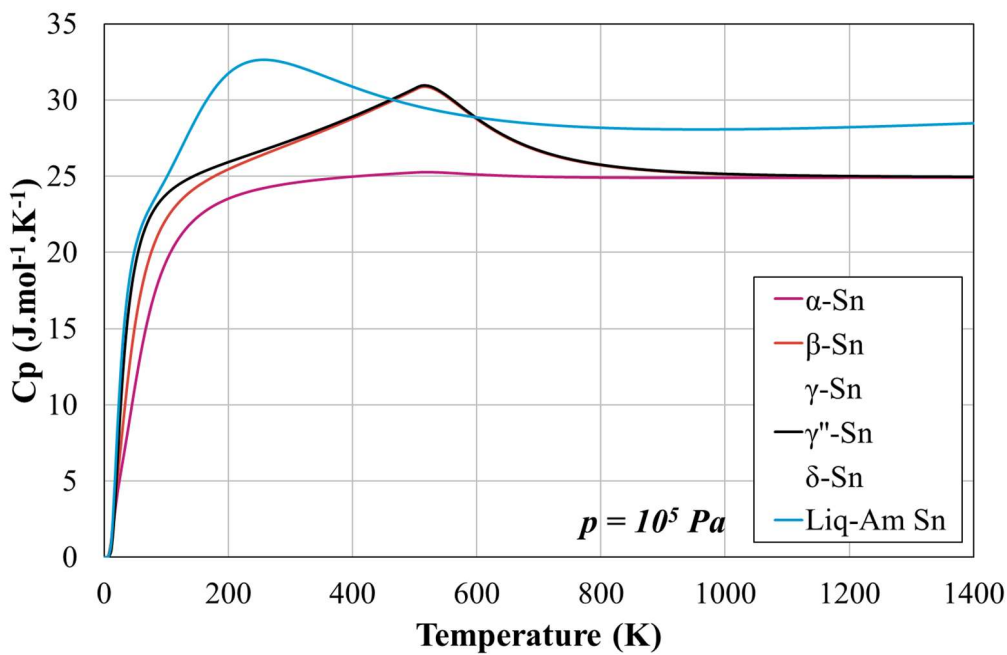

Fig. S4 – Heat capacity at atmospheric pressure of the phases modeled in this work. The heat capacity of the γ, γ'' and δ phases is too close to be distinguished.

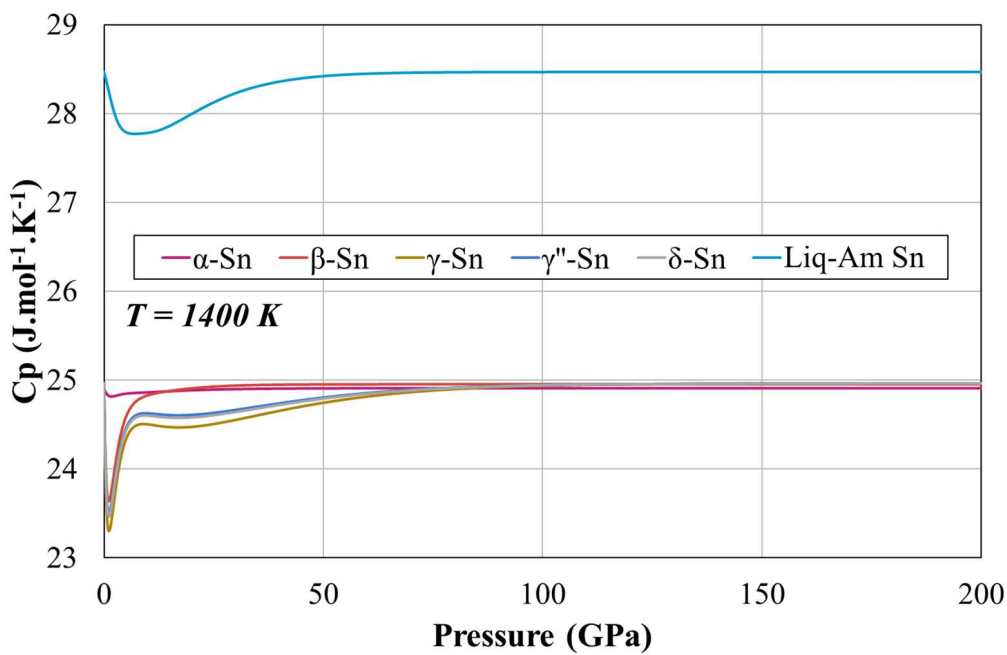

Fig. S5 – Heat capacity along the 1400 K isotherm of the phases modeled in this work. An approximate expression is used to extend the atmospheric pressure Gibbs energy toward high pressure, which result in the $C_p$ increasing back up to its $10^5$ Pa value at high pressure.

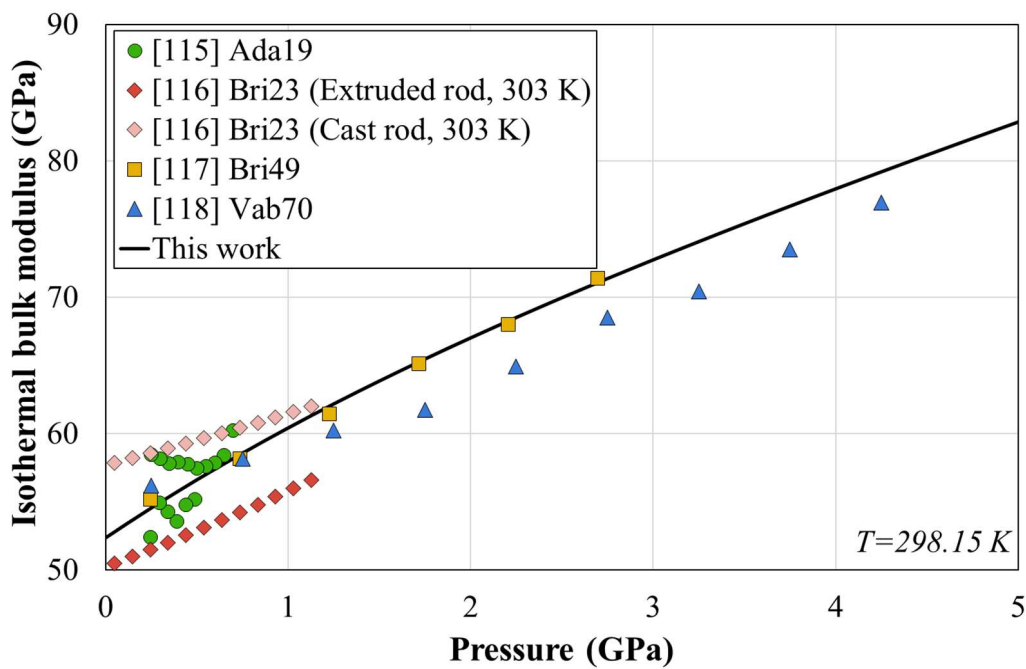

Fig. S6 – Bulk modulus of β-Sn calculated along the room temperature isotherm compared with experimental data

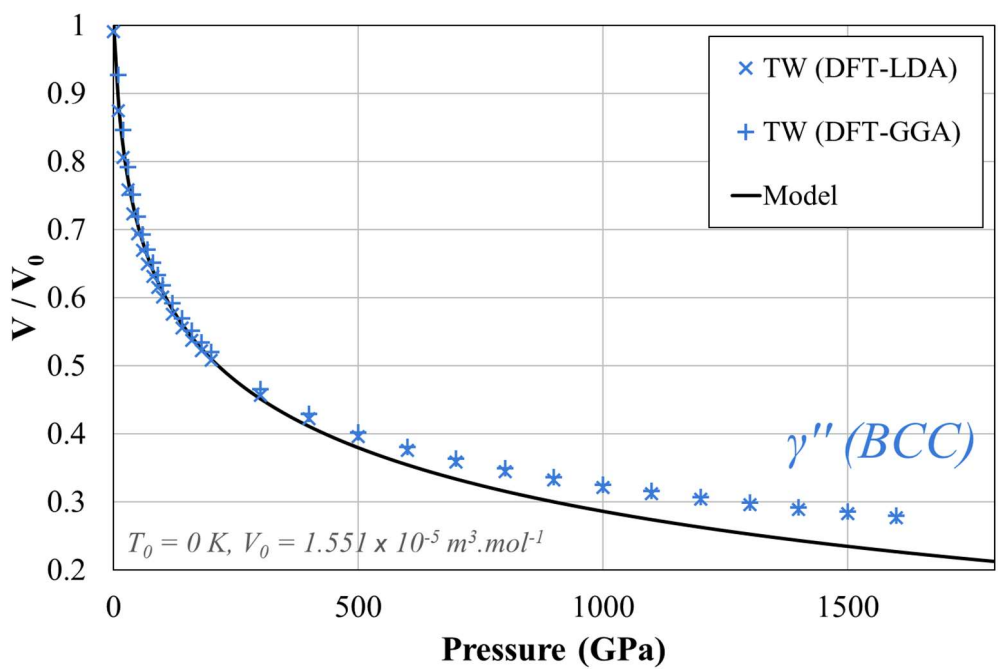

Fig. S7 – Volume of BCC Sn calculated along the 0 K isotherm compared with DFT calculations. Above pressures of twice the standard bulk modulus of the phase, the Grover empirical law [169] becomes less precise, and the volume starts to become underestimated.